\documentclass[12pt]{JHEP3}
\usepackage{mathrsfs}
\usepackage{amsmath,amssymb}
\usepackage{epsfig}
\input epsf

\def\x'{\mathaccent 19 x}
\def\y'{\mathaccent 19 y}
\def\n'{\mathaccent 19 n}
\def\u'{\mathaccent 19 u}

\def\et'{\mathaccent 19 \eta}
\def\th'{\mathaccent 19 \theta}
\def\lam'{\mathaccent 19 \lambda}
\def\varet'{\mathaccent 19 \vartheta}
\def\rh'{\mathaccent 19 \rho}
\def\ph'{\mathaccent 19 \phi}
\def\xb'{\mathaccent 19 {\bar{x}}}

\def\l{{\lambda}}

\def\N{${\cal N}=4$ }

\def \A {{\bf{A}}}

\def\det{\hbox{det}}
\def\be{\begin{equation}}
\def\ee{\end{equation}}

\newcommand{\bea}{\begin{eqnarray}}
\newcommand{\eea}{\end{eqnarray}}

\def\a {\alpha}
\def\b {\beta}
\def\s {\sigma}
\def\pa {\partial}
\def\P {\mathscr{P}}

\newcommand{\alg}[1]{\mathfrak{#1}}
\newcommand{\su}{\alg{su}}
\newcommand{\sls}{\alg{sl}}
\newcommand{\psu}{\alg{psu}}
\newcommand{\un}{\alg{u}}

\newcommand{\AdS}{{\rm  AdS}_5\times {\rm S}^5}

\def\L{\mathscr L}

\newcommand{\sfrac}[2]{{\textstyle\frac{#1}{#2}}}
\newcommand{\half}{\sfrac{1}{2}}
\newcommand{\ihalf}{\sfrac{i}{2}}

\preprint{ {\tt hep-th/0508140}\\ {\tt ITP-05/24} \\ {\tt
SPIN-05/30}\\
{\tt AEI-2005-136}}

\title{New Integrable System of 2dim Fermions from Strings on ${\bf \rm AdS}_5
\times {\bf \rm S}^5$}

\author{L. F. Alday$^{a}$\footnote{e-mail: L.F.Alday@phys.uu.nl, G.Arutyunov@phys.uu.nl,
frolovs@aei.mpg.de}, G. Arutyunov$^{a}$\footnote{Also at Steklov
Mathematical Institute, Moscow } and S. Frolov$^{b}$\footnote{
Also at SUNYIT, Utica, USA, and Steklov Mathematical Institute,
Moscow }
\\
\\
$^{a}$ {\it Institute for Theoretical Physics and Spinoza Institute, Utrecht University \\
~~3508 TD Utrecht, The Netherlands}\\
$^{b}$ {\it Max-Planck-Institut f\"ur Gravitationsphysik,
Albert-Einstein-Institut}\\
~~Am M\"uhlenberg 1, D-14476 Potsdam, Germany\\
}

\abstract{We consider classical superstrings propagating on $\AdS$
space-time. We consistently truncate the superstring equations of
motion to the so-called $\su(1|1)$ sector. By fixing the uniform
gauge we show that physical excitations in this sector are
described by two complex fermionic degrees of freedom and we
obtain the corresponding Lagrangian. Remarkably, this Lagrangian
can be cast in a two-dimensional Lorentz-invariant form. The
kinetic part of the Lagrangian induces a non-trivial Poisson
structure while the Hamiltonian is just the one of the massive
Dirac fermion. We find a change of variables which brings the
Poisson structure to the canonical form but makes the Hamiltonian
nontrivial. The Hamiltonian is derived as an exact function of two
parameters: the total ${\rm S}^5$ angular momentum $J$ and string
tension $\lambda$; it is a polynomial in $1/J$ and in
$\sqrt{\lambda'}$ where $\lambda'=\frac{\lambda}{J^2}$ is the
effective BMN coupling. We identify the string states dual to the
gauge theory operators from the closed $\su(1|1)$ sector of ${\cal
N}=4$ SYM and show that the corresponding near-plane wave energy
shift computed from our Hamiltonian perfectly agrees with that
recently found in the literature. Finally we show that the
Hamiltonian is integrable by explicitly constructing the
corresponding Lax representation.

}

\begin{document}

\newpage

\renewcommand{\thefootnote}{\arabic{footnote}}
\setcounter{footnote}{0}
\section{Introduction}
Further progress in understanding the AdS/CFT duality \cite{M} in
the large $N$ limit requires quantizing superstring theory on
$\AdS$. Even though classical superstring on $\AdS$ is an
integrable model \cite{Bena:2003wd} it is difficult to quantize it
by conventional methods developed in the theory of quantum
integrable systems \cite{FTa}. Action variables are encoded in
algebraic curves describing finite-gap solutions of the string
sigma-model \cite{Kazakov:2004qf}, however, angle variables have
not been yet identified.

 \medskip

On the other hand, the dilatation operator of ${\cal N}=4$ SYM can
be viewed as a Hamiltonian of an integrable spin chain
\cite{Minahan:2002ve} which at higher loops becomes long-range
\cite{Beisert:2003tq}. Perturbative scaling dimensions of
composite operators can be computed by solving the corresponding
Bethe ansatz equations \cite{BDS,BS2} \footnote{Related aspects of
integrability of strings on $\AdS$ and its gauge theory
counterpart were also studied in
\cite{Vallilo:2003nx}-\cite{Agarwal:2005ed} and subsequent
works.}.

\medskip

The success of the Bethe ansatz approach in gauge theory hints
that the spectrum of quantum strings might also be encoded in a
similar set of equations. Indeed, a Bethe type ansatz which
captures dynamics of quantum strings in certain asymptotic regimes
has been proposed \cite{Arutyunov:2004vx}. The quantum string
Bethe ansatz (QSBA) describes the spectrum of string states dual
to gauge theory operators from the closed $\su(2)$ sector
\cite{Arutyunov:2004vx}. The dual gauge theory contains other
closed sectors \cite{Beisert:2004ry}, and it is possible to
generalize the QSBA to these \cite{S}, and even to the complete
model \cite{BS2}. However, it remains unclear how the QSBA can
emerge from an exact (non-semiclassical) quantization of strings.

\medskip

It turns out that classical superstring theory on ${\rm
AdS}_5\times {\rm S}^5$ admits consistent truncations to smaller
sectors \cite{Alday:2005gi} which contain string states dual to
operators from the closed sectors of gauge theory. Apparently, the
truncated models are non-critical, and therefore, are expected to
loose many important features of the superstring theory on $\AdS$
such as conformal invariance and renormalizability. However, they
inherit classical integrability of the parent theory, and one
might hope that despite their apparent non-renormalizability there
would exist a unique quantum deformation which preserves
integrability and describes correctly the dynamics of quantum
superstrings in these sectors.

\medskip

As is known \cite{Beisert:2004ry}, \N SYM contains three simple
closed sectors: $\su(2)$, $\sls(2)$ and $\su(1|1)$. In the full
theory they are related to each other by supersymmetry which
implies highly nontrivial relations between the spectra of
operators from these sectors \cite{S}. The consistent truncations
of {\it classical} superstring theory to the $\su(2)$ and
$\sls(2)$ sectors  describe strings propagating in ${\mathbb
R}\times {\rm S}^3$ and ${\rm AdS}_3\times {\rm S}^1$,
respectively. A truncation of superstrings to the $\su(1|1)$
sector is unknown, and finding it is one of the aims of our paper.

\medskip

The $\su(1|1)$ sector of the gauge theory seems to be the simplest
one, in particular, the one-loop dilatation operator describes a
free lattice fermion \cite{Callan:2004dt}. In truncated string
theory one expects physical excitations to be carried by two
complex fermions, and, therefore, one might hope to find an action
which is polynomial in the fermionic variables. This would
represent a drastic simplification in comparison to the reductions
of superstrings to the $\su(2)$ and $\sls(2)$ sectors where
physical excitations are bosonic and described by non-polynomial
Nambu-type actions \cite{Arutyunov:2004yx}. Thus, finding quantum
deformations in the $\su(1|1)$ sector might be more feasible.

\medskip

Independently of the importance of this problem to the AdS/CFT
correspondence, finding consistent reductions of the superstring
theory provides a way to generate new interesting integrable
models. The simplest example of such a kind is the Neumann model
\cite{AFRT} describing rigid multi-spin string solitons \cite{FT}.
Among other examples of new integrable systems is the Nambu-type
Hamiltonian for physical degrees of freedom of bosonic strings on
$\AdS$ \cite{Arutyunov:2004yx}.

\medskip

We start from the classical string action on $\AdS$
\cite{Metsaev:1998it,Roiban} formulated as a sigma-model on the coset
PSU(2,2$|$4)/SO(4,1)$\times$SO(5). It is essential for our
approach to parametrize a coset representative by coordinates on
which the global symmetry group PSU(2,2$|$4) is linearly realized.
This makes the identification between these string coordinates and
the fields of the dual gauge theory transparent. It also allows us
to find easily a consistent truncation of the string equations of
motion to the $\su(1|1)$ sector. This procedure involves imposing
a so-called uniform gauge \cite{kt,Arutyunov:2004yx} which amounts
to identifying the global AdS time with the world-sheet time
$\tau$ and fixing the momentum of an angle variable of ${\rm S}^5$
to be equal to the corresponding U(1) charge $J$. Before the gauge
fixing the string Lagrangian of the reduced model has two bosons
and two complex fermions, and inherits two linearly realized
supersymmetries from the parent theory. Imposing the gauge
completely removes all the bosons so that the physical excitations
are carried  only by the fermions while supersymmetries become
non-linearly realized. Quite surprisingly, the two complex
space-time fermions can be combined into a single Dirac fermion,
$\psi$, and the action can be cast into a manifestly
two-dimensional Lorentz-invariant form. Thus, the original
Green-Schwarz fermions which are world-sheet scalars transform
into world-sheet spinors. This reminds the relation between the
flat space light-cone formulations of the Green-Schwarz and NSR
superstrings. In addition the Lagrangian exhibits the usual U(1)
symmetry which is realized by a phase multiplication of the Dirac
fermion.

\medskip

The Hamiltonian we obtain coincides with that of the massive Dirac
fermion. However, the kinetic part of the Lagrangian induces a
non-trivial Poisson structure which we explicitly describe. The
Poisson bracket is ultra-local, and is an 8-th order polynomial in
 the fermion $\psi$ and its first derivative. Then, we
show that there is a change of variables which brings the Poisson
structure to the canonical form but makes the Hamiltonian
nontrivial. We find the Hamiltonian as an exact function of two
parameters: the total ${\rm S}^5$ angular momentum $J$ and string
tension $\lambda$. It appears to be a polynomial in $1/J$ and in
$\sqrt{\lambda'}$ where $\lambda'=\frac{\lambda}{J^2}$ is the
effective BMN coupling.

\medskip

We can also use our Hamiltonian to study the near-plane wave
corrections to the energy of the plane-wave states from the
$\su(1|1)$ sector. To this end we keep in the Hamiltonian terms up
to order $1/J$, and compute the energy shift by using the
first-order perturbation theory. The same correction has been
already found in \cite{Callan:2004ev,McLoughlin:2004dh} by using a
light-cone type gauge. The uniform gauge we adopt in our approach
is different and that makes a comparison of their Hamiltonian with
ours difficult. Nevertheless, we demonstrate that the energy of an
arbitrary $M$-impurity plane-wave state computed by using our
Hamiltonian is in a perfect agreement with the results by
\cite{Callan:2004ev,McLoughlin:2004dh}. Thus, at the order $1/J$
our Hamiltonian leads to equivalent dynamics. Let us also mention
that the coherent state description of the $\su(1|1)$ sector with
its further comparison to string theory was considered in
\cite{Hernandez:2004kr}.

\medskip

Finally, we show that the Lax representation of the full string
sigma-model \cite{Bena:2003wd} also admits a consistent reduction
to the $\su(1|1)$ sector. Thus, the Hamiltonian of the reduced
model is also integrable.

\medskip

The paper is organized as follows. In section 2 we recall the
necessary facts about the Lie superalgebra $\psu(2,2|4)$, and the
construction of the string sigma-model Lagrangian. We also discuss
our specific choice for the coset representative  as well as the
global symmetries of the model. In section 3 we identify the
consistent truncation to the $\su(1|1)$ sector and in section 4 we
obtain the corresponding Lagrangian. In section 5 the Hamiltonian
and the Poisson structure of the model are found. By redefining
the fermionic variables we transform in section 6 the Poisson
structure to the canonical form and compute the accompanying
Hamiltonian.  In section 7 the near-plane wave energy shift
is computed, and in section 8 the Lax representation for the reduced
model is studied. Finally, some technical details and the Poisson
structure of the reduced model are collected in five appendices.

\section{Superstring on $\AdS$ as the coset sigma-model}
Superstring propagating in the $\AdS$ space-time can be described
as the non-linear sigma-model whose target space is the following
coset \cite{Metsaev:1998it} {\footnotesize \bea \label{ce}
\frac{\rm PSU(2,2|4)}{{\rm SO(4,1)}\times {\rm SO(5)}} \, . \eea }
\noindent Here the supergroup $\rm PSU(2,2|4)$ with the Lie
algebra $\psu(2,2|4)$ is the isometry group of the $\AdS$
superspace. The string theory action is the sum of the non-linear
sigma-model action and of the topological Wess-Zumino term to
ensure $\kappa$-symmetry.

In what follows we need to introduce a suitable parametrization
for the coset element (\ref{ce}). We start by recalling several
basic facts about the corresponding Lie superalgebra.

\subsection{The superalgebra $\psu(2,2|4)$}
The  superalgebra $\su(2,2|4)$ is spanned by $8\times 8$ matrices
$M$ which can be written in terms of $4\times 4$ blocks as \bea
M=\left(
\begin{array}{cc}
  A & X \\
  Y & D
\end{array} \right)\, .
\eea These matrices are required to have vanishing supertrace
${\rm str}M={\rm tr}A-{\rm tr}D=0$ and to satisfy the following
reality condition \bea \label{real} HM+M^{\dagger}H=0\, . \eea For
our further purposes it is convenient to pick up the hermitian
matrix $H$ to be of the form \bea  H=\left(
\begin{array}{rr}
  \Sigma & 0 \\
  0 & -\mathbb{I}
\end{array} \right)\, ,
\eea where $\Sigma$ is the following matrix \bea \Sigma=\,
{\footnotesize\left(
\begin{array}{cccc}
  1 & 0 & 0 & 0 \\
  0 & 1 & 0 & 0 \\
   0 & 0 & -1 & 0 \\
   0 & 0 & 0 & -1
\end{array} \right)} \eea
and $\mathbb{I}$ denotes the identity matrix of the corresponding
dimension. The matrices $A$ and $D$ are even, and $X,Y$ are odd
(linearly depend on fermionic variables). The condition
(\ref{real}) implies that $A$ and $D$ span the subalgebras
$\un(2,2)$ and $\un(4)$ respectively, while $X$ and $Y$ are
related through $Y=X^{\dagger}\Sigma$.  The algebra $\su(2,2|4)$
also contains the $\un(1)$ generator $i\mathbb{I}$ as it obeys
eq.(\ref{real}) and has zero supertrace. Thus, the bosonic
subalgebra of $\su(2,2|4)$ is \bea \su(2,2)\oplus \su(4)\oplus
\un(1)\, . \eea The superalgebra $\psu(2,2|4)$ is defined as the
{\it quotient algebra} of $\su(2,2|4)$ over this $\un(1)$ factor;
it has no realization in terms of $8\times 8$ supermatrices.

\medskip

The superalgebra $\su(2,2|4)$ has a $\mathbb{Z}_4$ grading
$$
M=M^{(0)}\oplus M^{(1)}\oplus M^{(2)}\oplus M^{(3)}
$$
defined by the automorphism $M\to \Omega(M)$ with \bea
\label{Omega} \Omega(M)= \left(
\begin{array}{rr}
  K A^t K ~&~ -K Y^tK \\
   K X^t K ~&~ K D^t K
\end{array} \right)\, ,
 \eea
where we choose the $4\times 4$ matrix $K$ satisfying $K^2=-I$ to be
\begin{equation}
K={\scriptsize\left(
\begin{array}{cccc}
  0 & -1 & 0 & 0 \\
  1 & 0 & 0 & 0 \\
   0 & 0 & 0 & -1 \\
   0 & 0 & 1 & 0
\end{array} \right)\, .}
\end{equation}
The space $M^{(0)}$ is in fact the ${\rm so(4,1)}\times {\rm
so}(5)$ subalgebra, the subspaces $M^{(1,3)}$ contain odd
fermionic variables.

\medskip

The orthogonal complement $M^{(2)}$ of ${\rm so(4,1)}\times {\rm
so}(5)$ in $\su(2,2)\oplus \su(4)$ can be conveniently described
as follows. In appendix A we introduce the matrices $\gamma_a$ and
$\Gamma_{a}$,  $a=1,\ldots , 5$, which are the Dirac matrices for
SO(4,1) and SO(5) correspondingly. These matrices obey the
relations \bea K\gamma_a^t K=-\gamma_a\, , ~~~~~ K\Gamma_{a}^t K
=-\Gamma_{a}\, \eea and, therefore, they span the orthogonal
complements to the Lie algebras so(4,1) and so(5) respectively.

\subsection{The Lagrangian}

Consider now a group element $g$ belonging to ${\rm PSU}(2,2|4)$
and construct the following current \bea \label{la} \A=-g^{-1}{\rm
d}g=\underbrace{\A^{(0)}+\A^{(2)}}_{\rm even
}+\underbrace{\A^{(1)}+\A^{(3)}}_{\rm odd}\, . \eea Here we also
exhibited the $\mathbb{Z}_4$ decomposition of the current. By
construction this current has zero-curvature.

The Lagrangian density for superstring on $\AdS$ can be written in
the form
\cite{Metsaev:1998it,Roiban}
\bea
\label{sLag} \L
=-\sfrac{1}{2}\sqrt{\lambda}\gamma^{\a\b}{\rm
str}\Big(\A^{(2)}_{\a}\A^{(2)}_{\b}\Big)-\kappa
\epsilon^{\a\beta}{\rm
str}\Big(\A^{(1)}_{\a}\A^{(3)}_{\beta}\Big)\, , \eea which is the
sum of the kinetic and the Wess-Zumino terms. $\kappa$-symmetry
requires $\kappa=\pm\sfrac{1}{2}\sqrt{\lambda}$. Here we use the
convention $\epsilon^{\tau\sigma}=1$ and $\gamma^{\a\b}= h^{\a\b}
\sqrt {-h}$ is the Weyl-invariant combination of the metric on the
string world-sheet with $\det\gamma=-1$.

\subsection{Coset Representative}
Obviously there are many different ways to parametrize the coset
element (\ref{ce}), all of them related by non-linear field
redefinitions. In what follows we find convenient to use the
following parametrization for the coset element \bea
g=g(\theta,\eta)g(x,y)\, . \eea Here $g(x,y)$ describes an
embedding of $\AdS$ into  $\mbox{SU(2,2)}\times \mbox{SU(4)}$ and
$g(\theta,\eta)$ is a matrix which incorporates the original 32
fermionic degrees of freedom. We take \bea \label{bc}
g(x,y)=\underbrace{\exp\half(x_{a}\gamma_a)}_{g(x)}
\underbrace{\exp\ihalf(y_a\Gamma_a)}_{g(y)} \eea Here the
coordinates $x_a$ parametrize the $\mbox{AdS}_5$ space while $y_a$
stand for coordinates of the five-sphere.
It is also understood that $g(x,y)$ is a 8 by 8 block-diagonal matrix with
the upper 4 by 4 block equal to $g(x)$, and the lower block equal
to $g(y)$.

Finally, the odd matrix is of the form (distinction between
$\theta$'s and $\eta$'s will be discussed later)
\begin{equation}
\label{fermb} g(\theta,\eta)=\exp{\scriptsize \left(
\begin{array}{cccccccc}
  0 & 0 & 0 & 0 & \eta^5 & \eta^6 & \eta^7 & \eta^8 \\
  0 & 0 & 0 & 0 & \eta^1 & \eta^2 & \eta^3 & \eta^4 \\
  0 & 0 & 0 & 0 & \theta^1 & \theta^2 & \theta^3 & \theta^4 \\
  0 & 0 & 0 & 0 &  \theta^5 & \theta^6 & \theta^7 & \theta^8 \\
 \eta_5 & \eta_1 & -\theta_1 & -\theta_5  & 0 & 0 & 0 & 0 \\
 \eta_6 & \eta_2 & -\theta_2 & -\theta_6  & 0 & 0 & 0 & 0 \\
 \eta_7 & \eta_3 & -\theta_3 & -\theta_7  & 0 & 0 & 0 & 0 \\
 \eta_8 &\eta_4 &  -\theta_4 & -\theta_8  & 0 & 0 & 0 & 0
\end{array}\right)
}\, .
\end{equation}
Here $\theta^i$ and $\eta^i$ are $8+8$ complex fermions obeying
the following conjugation rule $\theta^{i~*}=\theta_i$ and
$\eta^{i~*}=\eta_i$. By construction the element $g$ and,
$g(\theta,\eta)$ in particular, belong to the supergroup
SU(2,2$|$4).

\medskip

It is worth emphasizing that the parametrization of the coset
element we choose is different from the one used by Metsaev and
Tseytlin \cite{Metsaev:1998it}, in particular we put the matrix
containing fermionic variables to the left from the bosonic coset
representative. As we will see such a form of the coset element
makes the transformation properties of fermions under the global
symmetry group transparent and will allow us to easily identify
the consistent truncation.

\medskip

The bosonic coset element (\ref{bc}) provides parametrization of
the $\AdS$ space in terms of $5+5$ unconstrained coordinates $x_a$
and $y_a$. It is however more convenient to work with the
constrained $6+6$ coordinates which describe the embeddings of the
${\rm AdS}_5$ and the five-sphere into $\mathbb{R}^{4,2}$ and
$\mathbb{R}^6$ respectively. The latter parametrization was
introduced in \cite{AFRT}. Here the AdS and the sphere
representatives, $g_a(v)$ and $g_s(u)$, are described by the
following matrices \bea \label{GG}
 g_{a}(v)&=&{\footnotesize\left(\begin{array}{rrrr}
0 ~&~ -iv_5-v_6 ~&~ v_1-iv_4 ~&~ -iv_2-v_3 \\
iv_5+v_6 ~&~ 0 ~&~ -iv_2+v_3 ~&~ v_1+iv_4 \\
-v_1+iv_4 ~&~ iv_2-v_3 ~&~ 0 ~&~ iv_5-v_6 \\
iv_2+v_3 ~&~ -v_1-iv_4 ~&~ -iv_5+v_6 ~&~ 0
\end{array}
\right) } \, , \\ g_{s}(u)&=& {\footnotesize
\left(\begin{array}{rrrr}
0 ~&~ -iu_5-u_6 ~&~ -iu_1-u_4 ~&~ -u_2+iu_3 \\
iu_5+u_6 ~&~ 0 ~&~ -u_2-iu_3 ~&~ -iu_1+u_4 \\
iu_1+u_4 ~&~ u_2+iu_3 ~&~ 0 ~&~ iu_5-u_6 \\
u_2-iu_3 ~&~ iu_1-u_4 ~&~ -i u_5+u_6 & 0
\end{array}
\right)} \, . \eea The new variables $u,v$ are constrained \bea
\nonumber
&&v_1^2+v_2^2+v_3^2+v_4^2-v_5^2-v_6^2=-1\\
\label{adssphere} && u_1^2+u_2^2+u_3^2+u_2^2+u_5^2+u_6^2=1 \eea
which guarantees that $g_a(v)$ and $g_s(u)$ belong to SU(2,2) and
SU(4) respectively. On the coordinates $(u,v)$ the conformal and
R-symmetry transformations act {\it linearly} which is not the
case for $(x,y)$.

It is not difficult to find the explicit relation between these
two different description of the coset space.
Taking into account that arbitrary coset elements $g_a(v)$ and $g_s(u)$ of
SU(2,2)/SO(4,1) and SU(4)/SO(5) respectively can be represented in the form
\bea \label{nvsl} g_a(v)= g(x)K g(x)^t \, ,~~~~~~~
  g_s(u)=g(y) K g(y)^t \, ,
\eea
where $g(x)$ and $g(y)$ are SU(2,2) and SU(4) matrices, and choosing
them to be given by (\ref{bc}),
we see
that
the following relations are satisfied
 \bea
\label{cofv1}
 x_{a}&=&\frac{|x|}{\sinh |x|}v_{a},
~~~~~~~~|x|={\rm arcosh} v_6 \, ,\\
\label{cofv2} y_{a}&=&\frac{|y|}{\sin |y|}u_{a},
~~~~~~~~~~|y|=\arccos u_6 \, .\eea Here also \bea \nonumber
|x|^2=x_1^2+x_2^2+x_3^2+x_4^2-x_5^2\, , ~~~~~~~~
|y|^2=y_1^2+y_2^2+y_3^2+y_4^2+y_5^2 \, .\eea

As was mentioned above, the coordinates $(u,v)$ are very
convenient because they transform linearly under the isometry
group. In the following we first determine the Lagrangian of the
theory in terms of the coset element (\ref{ce}) and then
substitute in the final result the change of variables $(x,y)\to
(u,v)$ according to eqs.(\ref{cofv1}), (\ref{cofv2}).

\subsection{Global Symmetries}
To identify a consistent truncation to the $\su(1|1)$ sector we
have to analyze the global symmetries in more detail. According to
the standard technique of non-linear realizations the isometry
group ${\rm PSU}(2,2|4)$ acts on the coset representative by
multiplication from the left \bea Gg=g'g_{c}\, . \eea Here $G\in
{\rm PSU}(2,2|4)$, $g$ and $g'$ are the coset representatives
before and after the group action and $g_c$ is a {\it compensating
transformation} from SO(4,1)$\times$ SO(5). We will need  only
infinitesimal transformations generated by the algebra
$\psu(2,2|4)$.

\vskip 0.5cm \noindent{\sl Conformal Transformations of Bosonic
Fields} \vskip 0.5cm

Consider first the bosonic AdS coset element $g(x)$. We note that
since a matrix $A\equiv \half x_a\gamma_a$ obeys the relation $K
A^t K=-A$ the element $g$ itself also obeys \bea \label{coset} K
g(x)^t K  =-g(x)\, . \eea This gives a nice way to describe this
coset. The coset element is just a matrix from SU(2,2) group
obeying an additional constraint (\ref{coset}). An infinitesimal
conformal transformation reads \bea \label{trans} \delta g(x)=\Phi
g(x)-g(x)\Phi_c \eea Here $\Phi$ is an arbitrary matrix from the
Lie algebra $\su(2,2)$; it plays the role of the parameter of an
infinitesimal conformal transformation. The matrix $\Phi_c$
belongs to so(4,1)$\subset \su(2,2)$ and, therefore, it obeys the
relation \bea \label{ct} K \Phi_c^t K=\Phi_c \eea The element
$\Phi_c$ is not independent but should be found for a given $\Phi$
by requiring that $\delta g(x)$ also belongs to the coset, in
other words, \bea K\delta g(x)^t K=-\delta g(x)\, . \eea This
equation allows one to find the compensating so(4,1)
transformation $\Phi_c\equiv \Phi_c(\Phi,g)$. Actually to
determine the transformation law for the variables $v$ the
compensating matrix $\Phi_c$ is not needed. Indeed, using the
formula (\ref{nvsl}) we obtain \bea \label{trga} \delta
g_a(v)=\Phi g_a(v)+g_a(v)\Phi^t\, , \eea where $\Phi_c$ decouples
due to eq.(\ref{ct}). The explicit form of the transformation
rules for the coordinates $v$ can be found in appendix B.

In what follows we will be interested in the form of $\Phi$ corresponding to
translations of the global AdS time coordinate. The corresponding generator
is identified with the dilatation operator. As is clear from eq.(\ref{adssphere}),
the global AdS time coordinate $t$ can be expressed through $v_5$ and $v_6$ as
 follows
$$e^{it}=i v_5+v_6\ .$$
Then the U(1) subgroup which
rotates only $v_5$ and $v_6$
corresponds to translations of $t$. The explicit form of $\Phi$ can be easily
found from the formulas in appendix B, and is given by
\bea
\label{dilop}
\Phi={\footnotesize
\xi\,\frac{i}{2}\left(\begin{array}{rrrr}
1 ~&~ 0 ~&~  0 ~&~  0 \\
0 ~&~ 1 ~&~ 0 ~&~  0 \\
 0~&~  0 ~&~ -1 ~&~  0 \\
 0 ~&~  0 ~&~ 0 & -1
\end{array}
\right) }\, .
\eea
The time coordinate $t$ is shifted by the transformation by $\xi$:
$t\to t'=t + \xi$, and one can easily see by using formulas from appendix A that
the dilatation operator that generates the shift is
$$
\Phi_t = \sfrac{1}{2}\gamma_5 \, .
$$

\medskip

For our further purposes it is useful to identify the ${\rm
so(4)}\subset \su(2,2)$ symmetry which linearly rotates
$v_1,\ldots ,v_4$ but does not affect $v_{5,6}$ directions. It is
induced by the following matrix \bea \label{spin} \Phi_{\rm
so(4)}={\footnotesize \left(\begin{array}{rrrr}
i\xi_1 ~&~  \alpha_1+i\beta_1 ~&~ 0 ~&~ 0 \\
- \alpha_1+i\beta_1 ~&~ -i\xi_1 ~&~ 0 ~&~ 0 \\
0 ~&~ 0 ~&~ i\xi_3 ~&~  \alpha_6+i\beta_6 \\
0 ~&~ 0 ~&~ - \alpha_6+i\beta_6 & -i\xi_3
\end{array}
\right)} \,  \eea that is a direct sum of two $\su(2)$'s.

\vskip 0.5cm \noindent{\sl R-symmetry Transformations of Bosonic
Fields} \vskip 0.5cm

A similar analysis goes for the action of the $\su(4)$ R-symmetry
transformations. There are several interesting U(1) subgroups of
the $\su(4)$ algebra. To identify them we notice that the form of
$g_s$ in eq.(\ref{GG}) suggests to introduce the following three
complex scalars \bea Z_1=u_4+iu_1\, ,~~~~Z_2=u_2+iu_3\,
,~~~~~Z_3=u_6+iu_5\, \eea Then from Appendices B and A we deduce
that the field $Z_3$ carries a unit charge under the following
$\un(1)$ of $\su(4)$ generated by the matrix \bea \label{phi3}
\Phi_3 = {\footnotesize \frac{1}{2}\left(\begin{array}{rrrr}
1 ~&~ 0 ~&~  0 ~&~  0 \\
0 ~&~ 1 ~&~ 0 ~&~  0 \\
 0~&~  0 ~&~ -1 ~&~  0 \\
 0 ~&~  0 ~&~ 0 & -1
\end{array}
\right) }=\sfrac{1}{2}\Gamma_5\,.
\eea
The fields $Z_1$ and $Z_2$ are neutral under this U(1) group.

In the same way we find that $Z_1$ carries a unit charge and $Z_2$
and $Z_3$ are neutral under the $\un(1)$ generated by
$$
\Phi_1 = {\footnotesize
\frac{1}{2}\left(\begin{array}{rrrr}
1 ~&~ 0 ~&~  0 ~&~  0 \\
0 ~&~ -1 ~&~ 0 ~&~  0 \\
 0~&~  0 ~&~ 1 ~&~  0 \\
 0 ~&~  0 ~&~ 0 & -1
\end{array}
\right) }
\, ,
$$
and $Z_2$ carries a unit charge and $Z_1$ and $Z_3$ are neutral
under the $\un(1)$ generated by
$$
\Phi_2 = {\footnotesize
\frac{1}{2}\left(\begin{array}{rrrr}
-1 ~&~ 0 ~&~  0 ~&~  0 \\
0 ~&~ 1 ~&~ 0 ~&~  0 \\
 0~&~  0 ~&~ 1 ~&~  0 \\
 0 ~&~  0 ~&~ 0 & -1
\end{array}
\right) }
\, .
$$
\vskip 0.3cm Further we note that the last two $\un(1)$'s are subalgebras of
so(4)=$\su(2)\times \su(2)$ symmetry algebra  which rotates only $u_1,\ldots u_4$,
and is embedded in $\su(4)$
as
\bea
\Phi_{\rm so(4)}={\footnotesize \left(\begin{array}{rrrr}
i\xi_1 ~&~ \alpha_1+i\beta_1 ~&~ 0 ~&~ 0 \\
-\alpha_1+i\beta_1 ~&~ -i\xi_1 ~&~ 0 ~&~ 0 \\
0 ~&~ 0 ~&~ i\xi_3 ~&~ \alpha_6+i\beta_6 \\
0 ~&~ 0 ~&~ -\alpha_6+i\beta_6 & -i\xi_3
\end{array}
\right)} \, .
\eea

\vskip 0.5cm
\noindent{\sl Conformal and R-symmetry
Transformations of Fermions} \vskip 0.5cm

Let us now determine the transformation rules for the fermionic
variables under conformal and R-symmetry transformations. To
simplify the notation we  denote $g(\theta,\eta)=\exp\Theta$ and
the bosonic coset element by $g$. Then the infinitezimal action of
the symmetry group on fermions can be deduced from the general
formula describing the variation of the coset element
$$
\delta_{\Phi}(e^{\Theta}g)=\Phi e^{\Theta}g  - e^{\Theta}g
\Phi_c\, ,
$$
where $\Phi_c$ is again a compensating transformation which
generically might depend on $\Phi,g$ and $\Theta$. Taking into
account the expression (\ref{trans}) we find the transformation
rule for fermionic variables \bea \label{ferm}
\delta_{\Phi}\Theta=[\Phi,\Theta] \,  . \eea This shows an
advantage of our coset parametrization: the symmetries act
linearly on fermionic variables, just in the same manner as in the
dual gauge theory!

The similarity can be made even more explicit if we use (\ref{fermb})
to write the fermionic matrix $\Theta$ in the block form
$$
\Theta={\footnotesize \left(\begin{array}{rr}
0 ~&~ \widetilde{\Psi} \\
 \Psi~&~  0
\end{array}
\right)} \, ,
$$
and the conformal and R-symmetry transformations
matrix $\Phi$ in the block-diagonal
form
$$
\Phi={\footnotesize \left(\begin{array}{rr}
\Phi_a ~&~ 0 \\
 0~&~  \Phi_s
\end{array}
\right)} \, .
$$
Then, it is easy to see that
\bea
\label{ferm2}
\delta_{\Phi}\Psi= \Phi_s\,\Psi-\Psi\,\Phi_a\, ,\quad
\delta_{\Phi}\widetilde{\Psi}= \Phi_a\,
\widetilde{\Psi}-\widetilde{\Psi}\,\Phi_s \,.
\eea
It is clear from the formula that all columns of $\Psi$
transform in the fundamental
representation of $\su(4)$, and all
columns of $\widetilde{\Psi}$ transform in the fundamental
representation of $\su(2,2)$.

The transformation law (\ref{ferm2}) and the form of the dilatation matrix
(\ref{dilop}) can be used to determine that
all $\eta^i$ have charge $\sfrac{1}{2}$ under the
dilatation while the charge of $\theta^i$ is $-\sfrac{1}{2}$.
This explains
the notational distinction we made for the fermions $\eta$'s and
$\theta$'s.

\vskip 0.5cm \noindent{\sl Supersymmetry Transformations} \vskip
0.5cm

For the infinitezimal supersymmetry transformations with fermionic
parameter $\epsilon$ (comprising 32 supersymmetries) we find (up
to the linear order in $\Theta$ )
\bea
\label{ft}
\delta_{\epsilon}g&=&
\sfrac{1}{2}[\epsilon , \Theta ]g-g\Phi_c \, , \\
\delta_{\epsilon}\Theta&=&\epsilon \, .
\eea
Here again $g$ is the
bosonic coset element and  $\Phi_c\equiv \Phi_c(\epsilon,
\Omega)\in$so(4,1)$\times$so(5) should be determined from the
condition (\ref{coset}). For the elements $g_a(v)$ and $g_s(u)$
formula (\ref{ft}) implies \bea\nonumber
2\delta_{\epsilon}\left(\begin{array}{cc}
 g_a(v) & 0\\
0 &  g_s(u)
\end{array}\right)=[\epsilon , \Theta ]\left(\begin{array}{cc}
 g_a(v) & 0\\
0 & g_s(u)
\end{array}\right)+\left(\begin{array}{cc}
  g_a(v) & 0\\
0 &  g_s(u)
\end{array} \right)[\epsilon , \Theta ]^t\, .
 \eea
This concludes our discussion of the global symmetry
transformations.

\section{The $\su(1|1)$ sector of string theory}
We would like to find a consistent truncation of the superstring
equations to the smallest sector which should include the states
dual to the $\su(1|1)$ sector of the dual gauge theory. We
therefore start with recalling the necessary facts about the
$\su(1|1)$ sector of the gauge theory.

\medskip

The $\su(1|1)$ sector of ${\cal N}=4$ SYM comprises gauge
invariant  composite operators of the type
\bea
\label{gssu11}
{\rm tr}\big( \Psi^M Z^{J-\frac{M}{2}} \big)+\ldots
\eea
In the
${\cal N}=1$ language $Z$ stands for one of the three complex
scalar superfields, while $\Psi_{\a}$ is gaugino from the vector
multiplet. The field $\Psi_{\a}$ transforms as a spinor under one
of the $\su(2)$'s from the Lorentz algebra $\su(2,2)$ and is
neutral under the other. We use $\Psi$ to denote the highest
weight component of $\Psi_{\a}$. The fields $Z$ and $\Psi$ carry charges 1 and 1/2
under the U(1) subgroup of SU(4) generated by $\Phi_3$ (\ref{phi3}).
 By dots in eq.(\ref{gssu11}) we
mean all possible operators which can be obtained by permuting the
fermions inside the trace. In the free theory the conformal
dimension of the operators is $\Delta_0=J+M$ and the $\su(4)$
Dynkin labels $[0,J-\sfrac{M}{2},M]$.

\medskip

Coming back to string theory we notice that
the three complex
scalars $Z_i$ are naturally assumed to be dual to scalar
superfields of the gauge theory. Thus, reduction to the $\su(1|1)$
sector requires in particular to put, {\it e.g.}, $Z_1=Z_2=0$.
Also we put $v_1=\ldots=v_4=0$ leaving $v_{5,6}$ corresponding to
the global AdS time non-zero. The residual bosonic symmetry
algebra is then \bea {\rm so(4)}\times {\rm
so(4)}=\underbrace{\su(2)\times \su(2)}_{\rm AdS~~part} \times
\underbrace{\su(2) \times \su(2)}_{\rm sphere}\, . \eea

Taking into account eq.(\ref{ferm}) together with eq.(\ref{spin})
it is easy to see how the original 16 complex fermions are
decomposed w.r.t. the residual symmetry. Employing the notation of
\cite{Callan:2004ev} this decomposition can be described as
follows \bea (2,1;2,1)\oplus (2,1;1,2)\oplus
(1,2;2,1)\oplus(1,2;1,2) \, .\eea For instance an explicit form of
the fermionic matrix carrying irrep $(2,1;1,2)$ is
\begin{equation}
\Theta_{(2,1;1,2)}={\scriptsize \left(
\begin{array}{cccc|cccc}
  0 & 0 & 0 & 0 & 0 & 0 & \eta^7 & \eta^8 \\
  0 & 0 & 0 & 0 & 0 & 0 & \eta^3 & \eta^4 \\
  0 & 0 & 0 & 0 & 0 & 0 & 0 & 0 \\
  0 & 0 & 0 & 0 &  0 & 0 & 0 & 0 \\
  \hline
 0 & 0 & 0 & 0  & 0 & 0 & 0 & 0 \\
 0 & 0 & 0 & 0  & 0 & 0 & 0 & 0 \\
 \eta_7 & \eta_3 & 0 & 0  & 0 & 0 & 0 & 0 \\
 \eta_8 &\eta_4 &  0 & 0  & 0 & 0 & 0 & 0
\end{array}\right)
}\, .
\end{equation}
One can show that the string equations of motion admit a
consistent reduction to this sector which is governed by two
$\su(2)$-symmetries: one $\su(2)$ from the Lorentz algebra and
another one from the R-symmetry algebra, see also
\cite{Callan:2004ev}. However, this sector is still not the one to
put in correspondence with its counterpart from the dual gauge
theory.

As we will see a consistent truncation to a smaller set of
fermions exists. It amounts to putting
$$\eta^4=\eta^7=\eta_4=\eta_7=0$$
or, alternatively,
$$\eta^3=\eta^8=\eta_3=\eta_8=0\, .$$
To understand this we notice that in the ${\cal N}=1$ setting only
SU(3)$\times$U(1) subgroup of the R-symmetry group is manifest.
Since the gauge theory fermion $\Psi_{\a}$ belongs to the vector
multiplet it is neutral under $\su(3)$ which rotates three complex
scalars between themselves. On the string side the corresponding
Lie algebra element is
 \bea \Phi_{ \su(3)\times \underline{\un(1)} } ={\footnotesize
\left(\begin{array}{rrrr}
i\xi_1 ~&~ \alpha_1+i\beta_1 ~&~ \alpha_2+i\beta_2 ~&~ 0 \\
-\alpha_1+i\beta_1 ~&~ i\xi_2 ~&~ \alpha_4+i\beta_4 ~&~ 0 \\
-\alpha_2+i\beta_2 ~&~ -\alpha_4+i\beta_4 ~&~ i\xi_3 ~&~ 0 \\
0 ~&~ 0 ~&~ 0 & -i(\xi_1+\xi_2+\xi_3)
\end{array}
\right)} \, ,\eea where the $\su(3)$ part is specified by choosing
$\xi_3=-\xi_1-\xi_2$. Under the diagonal part of this matrix the
fields transform as
$$
\delta Z_1=i(\xi_1+\xi_3)Z_1\, , ~~~ \delta
Z_2=i(\xi_2+\xi_3)Z_2\, ,~~~ \delta Z_3=i(\xi_1+\xi_2)Z_3\, .
$$
and, therefore, the $\underline{\un(1)}$ part under which all
$Z$-fields carry the same charge is specified by
$\xi_1=\xi_2=\xi_3$. Again, by using eq.(\ref{ferm}) it is easy to
see that the fields which do not transform under $\su(3)$ are
$\eta^8$ and $\eta^4$. Thus, we would like to put one of these
fields in correspondence with the gauge theory fermion $\Psi$.
However, analyzing the structure of the cubic couplings in the
Lagrangian one can realize that the consistent reduction which
keeps non-zero only one fermion, say, $\eta^8$ {\it is not
allowed}. An obstruction arises due to the Wess-Zumino term as it
contains cubic couplings of two fermions with a single holomorphic
field $Z\equiv Z_3$. Switching off $\eta^4$ breaks the Lorentz
algebra $\su(2)$ down to $\un(1)$. Under this $\un(1)$ the field
$Z$ is uncharged while $\eta^8$ and $\eta^3$ carry opposite
charges, and therefore, they can form an invariant cubic coupling
of the type \bea \label{hc} e^{i t} Z\eta_3\eta_8 \eea with
possible $\tau$ and $\sigma$-derivatives acting on fermionic
fields. Moreover, one can easily see that this coupling is also
allowed by the $\underline{\un(1)}$-symmetry as the
$\underline{\un(1)}$ charge of $Z$ is precisely the opposite to
the sum of the fermionic charges. With the coupling eq.(\ref{hc})
in the Lagrangian it is inconsistent to put $\eta_3=0$ because the
equation of motion for $\eta_3$ will then turn into a non-linear
constraint involving $Z$ and $\eta_8$.

\medskip

Some comments are in order.   In what follows we will loosely
refer to the reduction which keeps only two complex non-zero
fermions, $\eta^3$ and $\eta^8$, as to the ``$\su(1|1)$ sector" of
string theory. The discussion above clearly demonstrates, however,
that consistent reductions of string equations of motion are not
the same as closed subsectors of the dual gauge theory as they
already contain different number of degrees of freedom. We will
return to the question about the relation between gauge and string
theory sectors in section 7. In this section we have identified a
possible truncation but we have not yet proven its consistency.
This will be done in section 8.

\section{Lagrangian of the reduced model}
In this section we fix a so-called uniform gauge
\cite{Arutyunov:2004yx} and derive the corresponding Hamiltonian.
In the uniform gauge approach the reparametrization freedom is
used to identify the world-sheet time $\tau$ with the global AdS
time $t$ and also to distribute a single component $J=J_3$ of the
${\rm S}^5$ angular momentum homogeneously along the string. We
will obtain the gauged-fixed Hamiltonian as an exact function of
$J$ and further show that in the large $J$ expansion it reproduces
the plane-wave Hamiltonian and higher-order corrections to it.

\medskip

In our reduction we choose to keep two complex fermions, $\eta_3$
and $\eta_8$. Let us recall the definitions
$$
iv_5+v_6=e^{it}\, ,~~~~~~Z=iu_5+u_6=e^{i\phi} \, .
$$
Here $t$ is the global AdS time and $\phi$ is an angle variable
along one of the ${\rm S}^1$'s embedded in ${\rm S}^5$. Since we
consider closed strings the embedding fields are assumed to be
periodic functions of $0\leq \sigma\leq 2\pi$. Periodicity then
implies \bea \label{period}\phi(2\pi)-\phi(0)=2\pi m,
~~~~m\in\mathbb{Z}\, . \eea The winding number $m$ describes how
many times the string winds around the circle parametrized by
$\phi$.

\medskip

Substituting the reduction into the Lagrangian (\ref{sLag}) we
find the following result
\bea
\L
&=&\sfrac{\sqrt{\lambda}}{2}\gamma^{\tau\tau}\left(\dot{t}^2-
\dot{\phi}^2+\sfrac{i}{2}(\dot{t}+\dot{\phi})
\zeta_{\tau}
\right) \\
\nonumber
&+&\sfrac{\sqrt{\lambda}}{2}\gamma^{\sigma\sigma}
\left({t'}^2-{\phi'}^2+\sfrac{i}{2}({t'}+{\phi'})
\zeta_{\sigma}
\right)\\
\nonumber
&+&\sqrt{\lambda}\gamma^{\tau\sigma}\left(\dot{t}t'-\dot{\phi}\phi'
+\sfrac{i}{4}(\dot{t}+\dot{\phi})
\zeta_{\sigma}
+\sfrac{i}{4}(t'+\phi')
\zeta_{\tau} \right) + \L_{\rm wz}\, .
\eea
Here and below dot and
prime denote the derivatives w.r.t. $\tau$ and $\sigma$
respectively. We have also introduced the following concise
notation for the fermionic contributions
\bea
\nonumber
\zeta_{\tau}&=&
\eta_i\dot{\eta}^i+ \eta^i\dot{\eta}_i\, ,\\
\label{constr}\zeta_{\sigma}&=& \eta_i{\eta'}^i+ \eta^i{\eta'}_i
\, , \eea where $i=3,8$. Inspection of the Wess-Zumino term shows
that it contains exponential terms of the type $e^{i(t+\phi)}$
which are absent in the kinetic part of the Lagrangian. All these
exponential terms can be removed by making the following rescaling
of fermions\footnote{The new rescaled fermions $\vartheta_{3,8}$
should not be mistaken with the original fermions $\theta_i$ in
$\Theta$ that were set to zero to get the $\su(1|1)$ sector.}
$$
\eta_{3,8}= e^{-\sfrac{i}{2}(t+\phi)} \vartheta_{3,8}\, ,~~~~~~~
\eta^{3,8}= e^{\sfrac{i}{2}(t+\phi)} \vartheta^{3,8}\, .
$$
The original fermions $\eta$ were charged under the two $\un(1)$
symmetries that shift $t$ and $\phi$. The new variables
$\vartheta$ do not carry these charges any more, they appear to be
{\it neutral}. This fact will play an important role in
constructing physical states dual to the gauge theory operators
from $\su(1|1)$ sector.

It is worth mentioning that because of the field redefinition the fermions
$\vartheta$ are periodic if the winding number $m$ is even, and anti-periodic if $m$
is odd.\footnote{The same effect was also found in the analysis of the
spectrum of fluctuations around a
multi-spin circular string \cite{FT2}. }

\medskip

After the rescaling the Lagrangian becomes \bea \label{Lagrangian}
\L
&=&\sfrac{\sqrt{\lambda}}{2}\gamma^{\tau\tau}\left(\dot{t}^2-
\dot{\phi}^2+\sfrac{i}{2}(\dot{t}+\dot{\phi})
\zeta_{\tau} -\sfrac{1}{2}(\dot{t}+\dot{\phi})^2 \Lambda
\right) \\
\nonumber
&+&\sfrac{\sqrt{\lambda}}{2}\gamma^{\sigma\sigma}\left({t'}^2-
{\phi'}^2+\sfrac{i}{2}({t'}+{\phi'})
\zeta_{\sigma}
 -\sfrac{1}{2}(t'+\phi')^2 \Lambda
\right)\\
\nonumber
&+&\sqrt{\lambda}\gamma^{\tau\sigma}\left(\dot{t}t'-\dot{\phi}\phi'
+\sfrac{i}{4}(\dot{t}+\dot{\phi}) \zeta_{\sigma}
+\sfrac{i}{4}(t'+\phi') \zeta_{\tau}
-\sfrac{1}{2}(\dot{t}+\dot{\phi})(t'+\phi')\Lambda \right) +
\L_{\rm wz}\, ,\eea where the Wess-Zumino term has a remarkably
simple form\footnote{We have also omitted a unessential total
derivative contribution.} \bea \label{WZ} \L_{\rm
wz}&=\frac{\kappa}{2}\Omega_{\tau}(t'+\phi')-\frac{\kappa}{2}
\Omega_{\sigma}(\dot{t}+\dot{\phi})
 \, .
 \eea
Here for various fermionic contributions we use the concise
notations \bea \label{constr1}
\begin{array}{lll}
 \zeta_{\tau}=\vartheta_i\dot{\vartheta}^i+\vartheta^i\dot{\vartheta}_i \, ,
 ~~&~~
\Omega_{\tau}=\vartheta_3\dot{\vartheta}_8+\vartheta_8\dot{\vartheta}_3
-\vartheta^3\dot{\vartheta}^8-\vartheta^8\dot{\vartheta}^3\, ,
~~&~~
\Lambda=\vartheta_i\vartheta^i\, ,\\
\zeta_{\sigma}=
\vartheta_i{\vartheta'}^i+\vartheta^i{\vartheta'}_i \, , ~~&~~
\Omega_{\sigma}=\vartheta_3\vartheta'_8+\vartheta_8\vartheta'_3
-\vartheta^3\vartheta'^8-\vartheta^8\vartheta'^3 \, . ~~&~~
\end{array}
\eea

\medskip

To further understand the dynamics of our reduced model we have to
identify the true (physical) degrees of freedom. The most elegant
way to achieve this goal is to construct the Hamiltonian formulation
of the model. Let us denote by $p_t$ and $p_{\phi}$ the canonical
momenta for $t$ and $\phi$. Computing from eq.(\ref{Lagrangian})
the momenta $p_t$ and $p_{\phi}$ we recast our Lagrangian in the
phase space form
\bea
\nonumber
\L&=&p_t\dot{t}+p_{\phi}\dot{\phi}+\frac{i}{4}(p_t-p_{\phi})
\zeta_{\tau}+\sfrac{\kappa}{2}(t'+\phi')\Omega_{\tau}\\
\nonumber &-&\frac{1}{\gamma^{\tau\tau}\sqrt{\lambda}}
\Big[\frac{1}{4}\left(p_t-p_\phi\right)\Big(p_t(2+\Lambda)+p_{\phi}(2-\Lambda)
+ 2\kappa\Omega_{\sigma} \Big)\\\nonumber
&+&\frac{\l}{4}(t'+\phi')\Big(t'(2-\Lambda)-\phi'(2+\Lambda)+i\zeta_{\sigma}\Big)\Big]\\
\label{psL}
&+&\frac{\gamma^{\tau\sigma}}{\gamma^{\tau\tau}}
\Big[p_t t'+p_{\phi}\phi'+\frac{i}{4}(p_t-p_{\phi})
\zeta_{\sigma}+\frac{\kappa}{2}(t'+\phi')\Omega_{\sigma}\Big]
\, .
\eea
As is usual in string theory with two-dimensional
reparametrization invariance the components of the world-sheet
metric enter in the form of the Lagrangian multipliers.

The uniform gauge amounts to imposing the following two conditions
\cite{Arutyunov:2004yx} \bea \label{gauge} t=\tau \, ,
~~~~~~~~~p_{\phi}=J\, .\eea The equations of motion for the phase
space variables follow from eq.(\ref{psL}). Upon substitution of
the gauge conditions (\ref{gauge}) some of these equations turn
into constraints which we have to solve in order to find the true
dynamical degrees of freedom. Let us also note that we do not
introduce here the canonical momenta for the fermionic variables.
Fermions are not involved in our gauge choice and, therefore, can
be treated at the final stage when all the bosonic type
constraints have been already solved.

\medskip

Let us now describe the procedure for finding the physical
Hamiltonian in more detail. First varying w.r.t.
$\gamma^{\tau\sigma}$ we obtain an equation for $\phi'$: \bea
\label{phipo} \phi'=-i\frac{p_t-p_{\phi}}{4p_{\phi}+
2\kappa\Omega_{\sigma}}\zeta_{\sigma}\, . \eea Variation w.r.t. to
$\gamma^{\tau\tau}$ gives an equation which we solve for $p_t$. We
find two solutions: $ p_t=p_{\phi}$ and \bea \label{second}
p_t=-\frac{p_{\phi}(2-\Lambda)+2\kappa\Omega_{\sigma}}{2+\Lambda}\,
\, . \eea The variable $p_t$ conjugated to the global AdS time $t$
is nothing else as the density of the space-time energy of the
string: $p_t=-\mathcal{H}$. Indeed, since we fix $t=\tau$ the
Noether charge corresponding to the global time translations
should coincide with the Hamiltonian ${\rm H}$ for physical
degrees of freedom: \bea {\rm H}=\int_0^{2\pi}\frac{{\rm
d}\sigma}{2\pi}\mathcal{H}\, . \eea  We pick up the second
solution (\ref{second}) to proceed because it has the correct
bosonic limit: $p_t=-p_{\phi}$ that is $\mathcal{H}=J$. Thus, we
have determined the Hamiltonian density \bea {\mathcal
H}=\frac{J(2-\Lambda)+2\kappa\Omega_{\sigma}}{2+\Lambda}\, .\eea
Recalling the explicit expressions for $\Lambda$ and
$\Omega_{\sigma}$ we see that the Hamiltonian density  does not
contain the time derivatives of the fermionic fields. We postpone
further discussion of $\mathcal{H}$ till we find solution of all
the constraints.

\medskip

Substituting the solution for $p_t$ into eq.(\ref{phipo}) we
obtain \bea \label{ps} \phi'=\frac{i\zeta_{\sigma}}{2+\Lambda} \,
.\eea Integrating over $\sigma$ and taking into account
(\ref{period}) we obtain a constraint \bea \label{lm} {\mathcal
V}=i\int \frac{{\rm
d}\sigma}{2\pi}\frac{\zeta_{\sigma}}{2+\Lambda}=m\, .\eea This
constraint is the level-matching condition which we will impose on
physical states of the theory. Actually the field $\phi$ is
non-physical. Its evolution equation can be found from
(\ref{Lagrangian}) by varying w.r.t. $p_{\phi}$: \bea \label{pt}
\dot{\phi}=\frac{2-\Lambda+i\zeta_{\tau}}{2+\Lambda}\, . \eea
Equations (\ref{pt}) and (\ref{ps}) determine $\phi$ in terms of
fermionic variables. Thus, upon imposing gauge conditions and
solving the constraints we obtain that the physical degrees of
freedom in the sector we consider are carried by fermionic fields
only.

\medskip

Finally, equations of motion for $p_t$ and $\phi$ can be solved
for the world-sheet metric. We find the following
result\footnote{The metric component $\gamma^{\tau\s}$ is
determined up to an arbitrary function of $\tau$ which we have
chosen to be zero. This function plays the role of the Lagrangian
multiplier to the level-matching constraint, {\it c.f.} the
corresponding discussion in \cite{Arutyunov:2004yx}. } \bea
\label{metric1}
\gamma^{\tau\tau}&=&\frac{i}{2\sqrt{\lambda}}\frac{\lambda\zeta_{\s}^2+4(2J+
\kappa\Omega_{\sigma})^2}
{(\zeta_{\tau}-4i)(2J+\kappa\Omega_{\sigma})-\kappa\zeta_{\sigma}
\Omega_{\tau}}\, ,\\
\label{metric2}
\gamma^{\tau\s}&=&\frac{i}{2\sqrt{\lambda}}
\frac{\lambda\zeta_{\s}(\zeta_{\tau}-4i)+4\kappa(2J+\kappa\Omega_{\sigma})
\Omega_{\tau}}
{(\zeta_{\tau}-4i)(2J+\kappa\Omega_{\sigma})-\kappa
\zeta_{\sigma}\Omega_{\tau}}
\, . \eea Clearly, due to the grassmanian nature of the fermionic
variables these and all the other expressions we obtain are
polynomial in fermions.

\medskip

Now substituting solutions of the constraints into
(\ref{Lagrangian}) we obtain the following gauge-fixed Lagrangian
 \bea
 \label{Lagf}
\L=-J-\frac{iJ\zeta_{\tau}+2\kappa\Omega_{\sigma}-2J\Lambda}{2+\Lambda}
-\frac{i\kappa}{2}\frac{\zeta_{\tau}\Omega_{\sigma}-
\zeta_{\sigma}\Omega_{\tau}}{2+\Lambda}\,
. \eea This Lagrangian exhibits very interesting features which
will be discussed in the next section.

\section{Hamiltonian and Poisson Structure of Reduced Model}
First we introduce a two-component complex (Dirac) spinor $\psi$
by combining the fermions as \bea \label{wsf}
\psi=\left(\begin{array}{c} \vartheta_3
\\ \vartheta^8
\end{array}\right)\,
\eea and also define the following Dirac matrices
\bea \rho^{0}=\left(\begin{array}{cc} -1 & 0 \\
0 & 1
\end{array}\right)\, , ~~~~~~~~~~~~~
\rho^{1}=\left(\begin{array}{cc} 0 & i \\ i & 0
\end{array}\right)\, .
\eea These matrices satisfy the Clifford algebra with the flat
metric of the Minkowski signature. We also define the Dirac
conjugate spinor $\bar{\psi}=\psi^{\dagger}\rho^0$. By using
various fermionic identities collected in appendix C the
Lagrangian (\ref{Lagf}) can be written as \bea \nonumber
\L&=&-J-\frac{J}{2}\left(i\bar{\psi}\rho^0\pa_{0}\psi-
i\pa_{0}\bar{\psi}\rho^0\psi\right)
+i\kappa(\bar{\psi}\rho^1\pa_1\psi-\pa_1\bar{\psi}\rho^1
\psi)+J\bar{\psi}\psi\\
\nonumber
&+&\frac{J}{4}\left(i\bar{\psi}\rho^0\pa_{0}\psi-i\pa_{0}
\bar{\psi}\rho^0\psi\right)\bar{\psi}\psi-
\frac{i\kappa}{2}(\bar{\psi}\rho^1\pa_1\psi-\pa_1\bar{\psi}
\rho^1\psi)\bar{\psi}\psi-\frac{1}{2}J(\bar{\psi}\psi)^2\\
\label{quadm} &+&\frac{\kappa}{2}\epsilon^{\a\beta}(
\bar{\psi}\pa_{\a}\psi~\bar{\psi}\rho^5\pa_{\beta}\psi-
\pa_{\alpha}\bar{\psi}\psi~\pa_{\beta}\bar{\psi}\rho^5\psi
)
  +\frac{\kappa}{8}\epsilon^{\alpha\beta}(\bar{\psi}\psi)^2
\pa_{\a}\bar{\psi}\rho^5
\pa_{\beta}\psi \, ,
 \eea
where $\rho^5=\rho^0\rho^1$. Finally, we note that the Lagrangian
(\ref{quadm}) can be further simplified if we perform the
following change of variables \bea \label{fshift} \psi &\to&
\psi+\frac{1}{4}\psi (\bar{\psi}\psi)\, , ~~~~~~~~~~~~
\bar{\psi}\to \bar{\psi}+\frac{1}{4}\bar{\psi} (\bar{\psi}\psi)\,
. \eea Indeed, after this shift we obtain \bea \nonumber
\L&=&-J-\frac{J}{2}\left(i\bar{\psi}\rho^0\pa_{0}\psi-
i\pa_{0}\bar{\psi}\rho^0\psi\right)
+i\kappa(\bar{\psi}\rho^1\pa_1\psi-\pa_1\bar{\psi}\rho^1\psi)+
J\bar{\psi}\psi\\
\label{quadms} &+&\frac{\kappa}{2}\epsilon^{\a\beta}(
\bar{\psi}\pa_{\a}\psi~\bar{\psi}\rho^5\pa_{\beta}\psi-
\pa_{\alpha}\bar{\psi}\psi~\pa_{\beta}\bar{\psi}\rho^5\psi
)
  -\frac{\kappa}{4}\epsilon^{\alpha\beta}(\bar{\psi}\psi)^2
\pa_{\a}\bar{\psi}\rho^5
\pa_{\beta}\psi \, ,
 \eea
\noindent Clearly, if we  now rescale  the world-sheet variable
$\sigma$ as \bea \sigma\to -\frac{2\kappa}{J}\sigma
\label{rescale} \eea then the Lagrangian density acquires the form
\bea \label{lorentz}
\L&=&J\Big[-1-\frac{1}{2}\left(i\bar{\psi}\rho^{\a}\pa_{\a}
\psi-i\pa_{\a}\bar{\psi}\rho^{\a}\psi\right)
+\bar{\psi}\psi
\\
\nonumber
 &&~~~-\frac{1}{4}\epsilon^{\a\beta}(
\bar{\psi}\pa_{\a}\psi~\bar{\psi}\rho^5\pa_{\beta}\psi-
\pa_{\alpha}\bar{\psi}\psi~\pa_{\beta}\bar{\psi}\rho^5\psi
)
  +\frac{1}{8}\epsilon^{\alpha\beta}(\bar{\psi}\psi)^2
\pa_{\a}\bar{\psi}\rho^5
\pa_{\beta}\psi \Big]\,
 \eea
and it defines a  {\it Lorentz-invariant theory of the Dirac
fermion on the flat two-dimensional world-sheet}! Original
space-time fermions of the Green-Schwarz superstring are combined
into spinors of the two-dimensional world-sheet. This is very
similar to the well-known relation between the light-cone
formulations of the NSR and Green-Schwarz superstrings in the flat
space. Our Lagrangian is however non-linear and extends up to six
order in fermions. If we then combine the prefactor $J$ in
eq.(\ref{lorentz}) with the transformation of the measure ${\rm
d}\sigma\to -\frac{2\kappa}{J}{\rm d}\sigma$ under
eq.(\ref{rescale}) we see that rescaling (\ref{rescale}) is
equivalent to restoring the overall $\sqrt{\lambda}$ dependence of
the Lagrangian; the whole dependence on $J$ goes to the
integration bound: $0\leq -\sigma\leq \frac{\pi J}{\kappa}$.
Finally, we note that it would be interesting to understand if and
how to rewrite the Lagrangian above as the covariant theory of the
Dirac fermion but on the curved world-sheet with the metric
(\ref{metric1}), (\ref{metric2}). From now on we fix
$\kappa=\frac{\sqrt{\lambda}}{2}$.

\medskip
The Lagrangian (\ref{quadms}) is also invariant under the global
U(1) symmetry $\psi\to e^{i\epsilon}\psi$. In fact this symmetry
is nothing else but the U(1) part of the Lorentz SU(2) subgroup
left unbroken upon the reduction, {\it c.f.} the corresponding
discussion in the previous section. Computing the corresponding
Noether charge $Q$ we find
 \bea \label{u1} Q=J\int \frac{{\rm d}\sigma}{2\pi}
\Big(\bar{\psi}\rho^0\psi-i\frac{\sqrt{\lambda'}}{2}
\bar{\psi}\rho^0\psi
(\bar{\psi}\rho^1\pa_1\psi-\pa_1\bar{\psi}\rho^1\psi)\Big)\, .
\eea This symmetry will play a crucial role in constructing the
physical states dual to gauge theory operators from the $\su(1|1)$
sector.

\medskip

To simplify our further discussion of the Hamiltonian and Poisson
structure of the reduced model it is convenient to rescale the
fermions as $\psi\to \frac{1}{\sqrt{J}}\psi$. The Lagrangian
(\ref{quadms}) shows the following structure \bea
\label{rf}\L=\L_{\rm kin}-{\mathcal H}\, , \eea where the
Hamiltonian density ${\mathcal H}$ is of a very simple form \bea
\label{simple} {\mathcal
H}&=&J-i\frac{\sqrt{\lambda'}}{2}(\bar{\psi}\rho^1\pa_1
\psi-\pa_1\bar{\psi}\rho^1\psi)-\bar{\psi}\psi\,
, \eea {\it i.e.} it is just the Hamiltonian density for a massive
two-dimensional Dirac fermion.
  The kinetic term $\L_{\rm kin}$ contains time derivatives
and it is this term which defines the Poisson structure of the
model: \bea \label{quad1}
\L_{\rm kin}&=&-\frac{1}{2}\left(i\bar{\psi}\rho^0\pa_{0}
\psi-i\pa_{0}\bar{\psi}\rho^0\psi\right)\\
\nonumber &-&\frac{\sqrt{\lambda'}}{2J}(
\bar{\psi}\pa_{1}\psi~\bar{\psi}\rho^5\pa_{0}\psi
-\pa_{1}\bar{\psi}\psi~\pa_{0}\bar{\psi}\rho^5\psi )
-\frac{\sqrt{\lambda'}}{8J^2}\epsilon^{\alpha\beta}
(\bar{\psi}\psi)^2\pa_{\a}\bar{\psi}\rho^5
\pa_{\beta}\psi \, .
 \eea
Let us now explain how to find the corresponding Poisson bracket.
Obviously, the canonical momentum conjugate to $\psi$ does not
depend on $\dot{\psi}$ and, therefore, implies the (second-class)
constraints between the phase-space variables. The standard way to
determine the Poisson structure in this case is to construct the
corresponding Dirac bracket. We, however, will solve this problem
in a simpler but equivalent way. Indeed, the equations of motion
that follow from eq.(\ref{quadms}) can be schematically
represented as \bea \label{dm}
\Omega_{ij}\dot{\chi}_j=\frac{\delta {\rm H}}{\delta\chi_i}\, .
\eea Here the index $i$ runs from $1$ to $4$ and we introduced the
four-component fermion $\chi=(\psi_1,\psi_2,\psi_1^*,\psi_2^*)$.
Denote by $\Omega^{-1}$ the inverse matrix. Then, eq.(\ref{dm})
can be written as \bea \label{im}
\dot{\chi}_i=\Omega^{-1}_{ij}\frac{\delta{\rm
H}}{\delta\chi_j}\equiv \{{\rm H},\chi_i\}\, . \eea Clearly,
$\Omega^{-1}$ defines the Poisson tensor which we are interested
in. Thus, all what we need to do is to compute from $\L_{\rm kin}$
the $4\times 4$ matrix $\Omega$ and then to invert it. Performing
the corresponding computation we find the following Poisson
structure
 \bea
\{\psi_i(\sigma),\psi_j(\sigma')\}&=&
-i\frac{\sqrt{\lambda'}}{4J}(\psi_k\psi_l)'\delta_{ij}\epsilon_{kl}
\delta(\sigma-\sigma')+...\\
\{\psi_i(\sigma),\psi^*_j(\sigma')\}&=&\Big[i\delta^{ij}+
i\frac{\sqrt{\lambda'}}{2J}
(\epsilon_{ik}\delta_{jl}\psi'_{l}\psi_{k}^*+\epsilon_{jk}
\delta_{il}\psi_{k}\psi'^{*}_l)\Big]\delta(\sigma
-\sigma') +... \eea where $\epsilon_{12}=1$. The Poisson bracket
appears rather non-trivial, it extends up to the 8th order in
fermion $\psi$ and its derivative $\psi'$, we refer the reader to
appendix E where the complete expression for the bracket is
presented.

\section{Canonical Poisson structure and Hamiltonian}
In the previous section we formulated our dynamical system in such
a way that it has a rather simple Hamiltonian but a relatively
complicated Poisson structure. In this section we find a further
transformation of the fermionic variables which brings the Poisson
structure of the model to the canonical form. Of course, the prize
we pay for simplification of the Poisson brackets is that under
this transformation the Hamiltonian becomes rather non-trivial.
The key idea is to find such a non-linear redefinition of the
fermionic variables which transforms the kinetic term in
eq.(\ref{quad1}) to the canonical form \bea \label{quad2} \L_{\rm
kin}&=&-\frac{i}{2}\left(\bar{\psi}\rho^0\pa_{0}\psi-\pa_{0}
\bar{\psi}\rho^0\psi\right)\,
. \eea Indeed, the kinetic term (\ref{quad2}) implies the standard
symplectic structure
 \bea
\label{cb}
 \{\psi^*_{\a}(\sigma),\psi_{\beta}(\sigma')\}=i
\delta_{\alpha\beta}\delta(\sigma-\sigma')\, .\eea The proper
redefinition can be found order by order in powers of fermions.
For the sake of simplicity we omit the corresponding calculations
and refer the reader to appendix D, where we give the final and
explicit form of the required change of variables. Substituting the
found redefinition of the fermions, eqs.(\ref{directshift}), into
eq.(\ref{simple}) we obtain the following Hamiltonian
 \bea \nonumber {\mathcal
H}=J&-&i\frac{\sqrt{\lambda'}}{2}(\bar{\psi}\rho^1\pa_1\psi-
\pa_1\bar{\psi}\rho^1\psi)-\bar{\psi}\psi\\
\nonumber
&+&\frac{1}{J}\Big[\frac{\lambda'}{2}\Big((\bar{\psi}\pa_1\psi)^2
+(\pa_1\bar{\psi}\psi)^2\Big)-i\frac{\sqrt{\lambda'}}{2}
~\bar{\psi}\psi
(\bar{\psi}\rho^1\pa_1\psi-\pa_1\bar{\psi}\rho^1\psi)\Big]\\
\nonumber &+&\frac{1}{J^2}\Big[-i\frac{ \lambda'^{\frac{3}{2}}
}{8}(\bar{\psi}\psi)^2\Big(\pa_1\bar{\psi}\rho^1\pa_1^2\psi-\pa_1^2
\bar{\psi}\rho^1\pa_1\psi\Big)
-\frac{3\lambda'}{8}(\bar{\psi}\psi)^2~\pa_1\bar{\psi}\pa_1\psi
\\
\nonumber &~&~~~~~~~~~~~~~~~~~~~
+i\frac{\lambda'^{\frac{3}{2}} }{2}\bar{\psi}\psi(\bar{\psi}
\pa_1\psi-\pa_1\bar{\psi}\psi)
\pa_1\bar{\psi}\rho^1\pa_1\psi\Big]\\
\label{complicated}
&-&\frac{1}{J^3}\Big[\frac{\lambda'^2}{2}(\bar{\psi}\psi)^2
(\pa_1\bar{\psi}\pa_1\psi)^2\Big]\,
. \eea Thus, our dynamical system is described now by the
Hamiltonian (\ref{complicated}) supplied with the canonical
Poisson bracket (\ref{cb}). Therefore, in the following we will
refer to eq.(\ref{complicated}) as to the {\it canonical}
Hamiltonian.

\medskip

The expression (\ref{complicated}) provides the canonical
Hamiltonian of the consistently truncated $\su(1|1)$ subsector of
the classical superstring theory on $\AdS$. It was derived as an
exact function of $J$. We have rearranged the final result
(\ref{complicated}) in the form of the large $J$ expansion with
$\lambda'=\frac{\lambda}{J^2}$ kept fixed.\footnote{The
rearrangement of the $1/\sqrt{\l}$ expansion in the form of the
large $J$ expansion with $\lambda'$ fixed is a generic fact valid
also for the expansion around multi-spin string configurations
\cite{FT2}.}

 Thus, the first line in
eq.(\ref{complicated}) is the well-known plane-wave Hamiltonian
\cite{Metsaev:2001bj} and the second one encodes the near-plane
wave correction to it. It is rather intriguing  that $1/J$
expansion of $\mathcal{H}$ terminates at order $1/J^3$. This does
not happen, for instance, for the bosonic $\su(2)$ subsector of
string theory, where the uniform-gauge Hamiltonian is of the Nambu
(square root) type. Apriori one could expect the appearance of
higher derivative terms in eq.(\ref{complicated}) that would lead
to higher-order terms in $1/J$ (and also in $\lambda'$) expansion.
We believe that such a property of the $1/J$ expansion should have
certain implications for the dual gauge theory.

\medskip

To conclude this section we note that under redefinition
(\ref{directshift}) the level-matching constraint (\ref{lm})
becomes very simple \bea {\mathcal V}=\int \frac{{\rm d}\sigma}{2
\pi}
\frac{i}{2}(\bar{\psi}\rho^0\pa_1\psi-\pa_1\bar{\psi}\rho^0\psi)=i\int
\frac{{\rm d}\sigma}{2\pi} \psi_i^*\psi'_i\eea and it just
generates the rigid shifts $\sigma\to \epsilon\sigma$. Also the
generator Q of the U(1) charge (\ref{u1}) simplifies to \bea
Q=\int \frac{{\rm d}\sigma}{2\pi} \bar{\psi}\rho^0\psi=\int
\frac{{\rm d}\sigma}{2 \pi} \psi_i^*\psi_i\, . \eea This
simplification of the level-matching constraint and the U(1)
charge can be also considered as an independent non-trivial check
of redefinitions (\ref{directshift}).

\section{Near-plane wave correction to the energy}
The near-plane wave correction to the energy of the plane-wave
states from the $\su(1|1)$ sector has been already found in
\cite{Callan:2004ev,McLoughlin:2004dh}. The corresponding
computation  was based on finding the $1/J$ correction to the
plane-wave Hamiltonian in a specific light-cone type gauge. The
uniform gauge we adopt in our approach
 is different. Due to the
complicated nature of the results of \cite{Callan:2004ev} we were
not able to compare directly their Hamiltonian with the $1/J$-term
in eq.(\ref{complicated}). Moreover, we see that this comparison
will definitely require finding a redefinition of our fermionic
variables to that of \cite{Callan:2004ev}. Nevertheless it is
possible to make a comparison in a simple way. In this section we
compute the $1/J$ correction to the energy of arbitrary
$M$-impurity plane-wave states from our Hamiltonian
(\ref{complicated}) and find perfect agreement with the results in
\cite{Callan:2004ev,McLoughlin:2004dh}. The simplicity of the corresponding
calculation is rather remarkable.

\medskip

 To create string states dual to the gauge theory operators
from the $\su(1|1)$ subsector we need to choose a proper
representation of the anti-commutation relations for fermions.
Writing $\psi$ as \bea \psi=\left(\begin{array}{c} \psi_1 \\
\psi_2
\end{array}\right)\ ,
\eea and expanding the fermions in Fourier modes
\bea \psi_\a(\s)
= \sum_{n=-\infty}^\infty\, e^{i n \s}\psi_{\a,n}\ ,\qquad
\psi_\a^\dagger(\s) = \sum_{n=-\infty}^\infty\, e^{-i n
\s}\psi^\dagger_{\a,n}\ ,
\eea
we introduce the following creation
and annihilation operators
\bea \nonumber\psi_{1,n} &=&
f_na^+_n+g_n b_n^-\, ,~~~~~~~~~\psi_{2,n} =  f_nb^-_n+g_n
a_n^+\, ,\\
\psi^\dagger_{1,n}&=& f_na^-_n-g_n b_n^+\,
,~~~~~~~~~\psi^\dagger_{2,n}= f_nb^+_n-g_n a_n^-\, ,
\label{ferrep}
\eea
where we have
defined the functions \bea
\nonumber f_n=\sqrt{{1\over 2} + {1\over 2\sqrt{1 + \l' n^2}}}\, ,
~~~~~~~ g_n={i\sqrt{\l'}\, n\over 1+\sqrt{1 + \l'
n^2}}\,\sqrt{{1\over 2} + {1\over 2\sqrt{1 + \l' n^2}}} \, .\eea

 In terms of the oscillators,
the free Lagrangian which is the first line in eq.(\ref{quadm})
takes the form \bea \label{lagr} {\cal L} =-J+
\sum_{n=-\infty}^\infty\,\left[ -i \left( a^+_n\dot{a}^-_n
+b^+_n\dot{b}^-_n\right) - \omega_n\left( a^+_n a^-_n +b^+_n
b^-_n\right)\right] \, ,\eea where $\omega_n=\sqrt{1+\lambda'
n^2}$. We thus see that $(a^-, a^+)$ and $(b^-, b^+)$ are pairs of
canonically conjugated operators. The SYM operators from the
$\su(1|1)$ subsector are dual to states obtained by acting by
operators $a^+_n$ on the vacuum. In general, however, such a state
with $M$ excitations (``impurities''), $a^+_{n_1}\cdots a^+_{n_M}$
can be also multiplied by a function of $a^+_kb^+_m$ because the
combination $a^+_kb^+_m$ is neutral. It does not matter at the
first order in the $1/J$ expansion.

The level matching condition has the usual form \bea
\label{levmat} {\cal V} =
\frac{1}{J}\sum_{n=-\infty}^\infty\,\left(n\, a^+_n a^-_n - n\,
b^+_n b^-_n\right)\ , \eea and therefore the sum of $a$-modes
should be equal to the sum of $b$-modes. For the states dual to
SYM operators from the $\su(1|1)$ subsector the sum of modes
should vanish.

\medskip

Now we can compute the energy shift at order $1/J$. The relevant
part of the Hamiltonian (\ref{complicated}) is
 \bea \nonumber {\mathcal
H}&=&J-i\frac{\sqrt{\l'}}{2}(\bar{\psi}\rho^1\pa_1\psi-
\pa_1\bar{\psi}\rho^1\psi)-\bar{\psi}\psi\\
\nonumber &+&\frac{1}{J}\Big[\frac{\l'}{2}\Big((\bar{\psi}
\pa_1\psi)^2+(\pa_1\bar{\psi}\psi)^2\Big)-i\frac{\sqrt{\l'}}{2}
~\bar{\psi}\psi
(\bar{\psi}\rho^1\pa_1\psi-\pa_1\bar{\psi}\rho^1\psi)\Big] \,
.\eea We need to substitute here the representation for fermions,
eqs.(\ref{ferrep}), and switch off the $b$-oscillators. The
normal-ordered Hamiltonian is \bea \nonumber &&{\rm
H}=J+\sum_n\omega_n a^+_na^-_n
+\frac{\sqrt{\lambda'}}{2J}\sum_{n_1,n_2,n_3,n_4}
\delta_{n_1-n_2+n_3-n_4}(f_{n_1}f_{n_2}+g_{n_1}g_{n_2})\times \\
\nonumber &\times &\big[i(n_3+n_4)(f_{n_4}g_{n_3}+f_{n_3}g_{n_4})
-\sqrt{\lambda'}(n_1n_3+n_2n_4)
(f_{n_3}f_{n_4}+g_{n_3}g_{n_4})\big]~a^+_{n_4}a^+_{n_2}
a^-_{n_3}a^-_{n_1}\, . \eea

\noindent A state carrying $M$ units of the U(1) charge $Q$ is
\bea
\label{state}
|M\rangle =a_{n_{1}}^+\ldots a_{n_{M}}^+
|0\rangle \,.
\eea
Since all fermions $\psi_\a$ are neutral under
the U(1) subgroup rotating the bosonic field $Z$, any such a state
carries the same $J$ units of the corresponding charge for any
number of excitations $M$. That means that an $M$-impurity string
state should be dual to the field theory operator of the form
$$
{\rm tr}\big( \Psi^M Z^{J-\frac{M}{2}} \big)+\ldots \,.
$$
We can see from this formula that at $M=2J$
there should exist only one string state which is dual to the operator
$$
{\rm tr}\, \Psi^{2J}\,.
$$
Such a restriction cannot be seen in the $1/J$ perturbation theory
but would play an important role in the exact (finite $J$) quantization of the model.

\medskip

It is trivial to compute the matrix element
$$
\langle M|a^+_{n_4}a^+_{n_2} a^-_{n_3}a^-_{n_1}|M\rangle
=\frac{1}{2}\sum_{i,j=1
}^M\Big(\delta_{n_1,n_j}\delta_{n_3,n_i}-\delta_{n_1,n_i}\delta_{n_3,n_j}\Big)
\Big(\delta_{n_2,n_i}\delta_{n_4,n_j}-\delta_{n_2,n_j}\delta_{n_4,n_i}\Big)\,
,
$$
where $n_i$ and $n_j$ are some indices which occur in
(\ref{state}). With this formula at hand we can easily find the
energy shift ($\omega_i \equiv \omega_{n_i}$ ) \bea \langle M|{\rm
H}|M\rangle =J&+&\sum_{i=1}^M\omega_i -\frac{\lambda'}{4J}\sum_{i
\neq j}^M\frac{n_i^2+n_j^2+2n_i^2n_j^2
\lambda'-2n_in_j\omega_i\omega_j}{\omega_i\omega_j}\, . \eea  This
{\it precisely} reproduces
 the $1/J$ correction to the $M$-impurity
plane-wave states obtained in \cite{McLoughlin:2004dh}, which up
to order $\lambda'^2$ agrees with the gauge theory result
\cite{S}.

\section{Lax representation}
In this section we discuss the Lax representation of the equations
of motion corresponding to the truncated Lagrangian. Our starting
point is the Lax pair found in \cite{Bena:2003wd}. It is based on
the two-dimensional Lax connection $\mathscr{L}$ with components
 \bea
 \label{wL}
\mathscr{L}_{\a}=\ell_0 A_{\a}^{(0)}+\ell_1 A_{\a}^{(2)}
+\ell_2\gamma_{\a\b}\epsilon^{\beta\rho}A_{\rho}^{(2)} +\ell_3
Q_{\a}^++\ell_4 Q_{\a}^-\, , \eea where $\ell_i$ are constants and
$Q^{\pm}=A^{(1)}\pm A^{(3)}$.
 The connection $\mathscr{L}$ is required
to have zero curvature \bea \label{zc}
\pa_{\a}\mathscr{L}_{\beta}-\pa_{\b}\mathscr{L}_{\a}-
[\mathscr{L}_{\a},\mathscr{L}_{\b}]=0 \eea as a consequence of the
dynamical equations and the flatness of $A_{\a}$. This requirement
of zero curvature also leads to determination of the constants
$\ell_i$. First we find
$$\ell_0=1, ~~~~~~~~\ell_1=\frac{1+x^2}{1-x^2}\, ,$$
where $x$ is a {\it spectral parameter}. Then for the remaining
$\ell_i$ we obtain the following solution \bea \label{fp1}
\ell_2&=&s_1\frac{2x}{1-x^2}\, ,
~~~~~\ell_3=s_2\frac{1}{\sqrt{1-x^2}}\,
,~~~~~~~\ell_4=s_3\frac{x}{\sqrt{1-x^2}} \, . \eea Here
$s_2^2=s_3^2=1$ and \bea s_1+s_2s_3&=&0\, ~~~~~~~{\rm
for}~~~~~\kappa=\frac{\sqrt{\lambda}}{2}\, ,\\
s_1-s_2s_3&=&0\, ~~~~~~~{\rm
for}~~~~~\kappa=-\frac{\sqrt{\lambda}}{2}\, . \eea Thus, for every
choice of $\kappa$ we have four different solutions for $\ell_i$
specified by the choice of $s_2=\pm 1$ and $s_3=\pm 1$, {\it c.f.}
the corresponding discussion in \cite{Alday:2005gi}.  As explained
in \cite{Alday:2005gi}, the Lax connection (\ref{wL}) can be
explicitly realized in terms of $8\times 8$ supermatrices from the
Lie algebra $\su(2,2|4)$. In the algebra $\su(2,2|4)$ the
curvature (\ref{zc}) of $\L_{\a}$ is not exactly zero, rather it
is proportional to the identity matrix (anomaly) with a
coefficient depending on fermionic variables. However, in
$\psu(2,2|4)$ the curvature is regarded to be zero since
$\psu(2,2|4)$ is the factor-algebra of $\su(2,2|4)$ over its
central element proportional to the identity matrix
\cite{Alday:2005gi,Beisert:2005bm}. In the following we consider
the Lax connection which corresponds to the choice
$\kappa=\frac{\sqrt{\lambda}}{2}$.

\medskip

Now we are ready to show that the Lax connection (\ref{wL}) for
the general $\psu(2,2|4)$ model can be consistently reduced to a
Lax connection encoding the equations of motion of physical fields
from the $\su(1|1)$ sector. The fact that the reduction holds at
the level of the matrix equations formulated in terms of $8\times
8 $ matrices is rather non-trivial and should be regarded as a
proof of consistency of the reduction procedure.

\medskip

We start with the projection $A_{\a}^{(0)}$. As was already
discussed, in our reduction we keep non-zero only the Dirac
fermion $\psi$ and solve for the world-sheet metric
$\gamma^{\a\beta}$ and unphysical fields $t$, and $\phi$ in terms
of $\psi$ by using our uniform gauge conditions and the
constraints. Let us now compute the components $A_{\a}^{(0)}$ on
our reduction and further perform the shift (\ref{fshift}). We
find \bea
A_{\sigma}^{(0)}&=&\sfrac{1}{4}(1+\bar{\psi}\psi)
(\bar{\psi}\psi'-\bar{\psi}'\psi)~{\rm
diag} \Big(1,-1,0,0;0,0,-1,1\Big)\, ,
\nonumber\\
A_{\tau}^{(0)}&=&\Big[\sfrac{1}{4}(1+\bar{\psi}\psi)
(\bar{\psi}\dot{\psi}-\dot{\bar{\psi}}\psi)+\sfrac{i}{2}
\bar{\psi}\rho^0\psi \Big]~{\rm diag} \Big(1,-1,0,0
;0,0,-1,1\Big)\, . \nonumber \eea Thus, the component $A^{(0)}$
appears to be a diagonal matrix, the first (last) four eigenvalues
correspond to the AdS (sphere) part of the model. These matrices
have four zero's in the middle and this suggests that the whole
Lax connection for the reduced sector can be formulated in terms
of $4\times 4$ matrices rather than $8\times 8$. Computation of
the other components of the Lax connection shows that this is
indeed the case. Therefore, in what follows we compute the
components of the reduced Lax connection as traceless $8\times 8$
matrices and then throw away from all the matrices the $4\times 4$
block sitting in the middle ({\it i.e} the corresponding rows and
columns). This block appears to be non-trivial only for
$A_{\a}^{(2)}$, however, one can show that it leads to redundant
equations which are satisfied due to the equations of motions for
fermions followed from other matrix elements. To simplify our
treatment in what follows we present the reduced Lax connection in
terms of the $4\times 4$ matrices whose dynamical variables are
those from the Lagrangian (\ref{quadms}). It is convenient to
introduce two diagonal matrices \bea {\bf I}={\rm diag}
\big(1,-1,-1,1\big)\, , ~~~~~~{\bf J }={\rm diag}
\big(1,1,-1,-1\big)\, . \eea Then we find the following bosonic
currents for the reduced model \bea \nonumber
A_{\sigma}^{(0)}&=&\sfrac{1}{4}(1+\bar{\psi}\psi)
(\bar{\psi}\psi'-\bar{\psi}'\psi)~{\bf I} \, ,\\
\nonumber
A_{\tau}^{(0)}&=&\Big[\sfrac{1}{4}(1+\bar{\psi}\psi)
(\bar{\psi}\dot{\psi}-\dot{\bar{\psi}}\psi)+\sfrac{i}{2}\bar{\psi}\rho^0\psi
\Big]~{\bf I} \, ,\\
\nonumber
A_{\sigma}^{(2)}&=&\sfrac{1}{8}\zeta_{\s}~{\bf J}\, ,\\
\label{cur}
A_{\tau}^{(2)}&=&\sfrac{1}{8}(\zeta_{\tau}+2i\bar{\psi}\psi-4i)~{\bf
J}\, .
 \eea
Here for reader's convenience we recall that \bea
\zeta_{\tau}=\bar{\psi}\rho^0\dot{\psi}-\dot{\bar{\psi}}\rho^0\psi\,
,~~~~~~~~
\zeta_{\sigma}=\bar{\psi}\rho^0\psi'-\bar{\psi}'\rho^0\psi\, .
 \eea
Notice also that the coefficient of $A_{\sigma}^{(2)}$ is
proportional to the density of the level-matching condition. The
odd matrices $Q^{\pm}_{\a}$ appears on our reduction are precisely
skew-diagonal. Introducing the matrices
\begin{equation}
\label{ferm1} {\Theta}=(1+\sfrac{1}{4}\bar{\psi}{\psi}){\scriptsize
\left(
\begin{array}{cccc}
   &  &  & \psi_2 \\
     &  & \psi_1^* &  \\
  & \psi_1&  &     \\
 \psi_2^* &  &  &
\end{array}\right)
}\, , ~~~~~~~~
\hat{{\Theta}}=i(1+\sfrac{1}{4}\bar{\psi}{\psi}){\scriptsize
\left(
\begin{array}{cccc}
     &  &  & -\psi_1 \\
     &  & -\psi_2^* &  \\
  & \psi_2 &  &     \\
 \psi_1^* &  &  &
\end{array}\right)
}
\end{equation}
the components $Q_{\a}^{\pm}$ can be written as
\bea\nonumber
Q_{\a}^+=[A_{\a}^{(2)},\Theta]-\pa_{\a}\Theta \, ,~~~~~~~
Q_{\a}^-=[A_{\a}^{(2)},\hat{\Theta}]+\pa_{\a}\hat{\Theta} \, .
\eea

\medskip

\noindent The original Lax connection (\ref{wL}) also involves the
following terms
\bea \nonumber
&&\gamma_{\tau\b}\epsilon^{\beta\rho}A_{\rho}^{(2)}=
-\gamma^{\s\s}A_{\s}^{(2)}-\gamma^{\sigma\tau}A_{\tau}^{(2)}\, ,\\
\label{mi} &&
\gamma_{\sigma\b}\epsilon^{\beta\rho}A_{\rho}^{(2)}=
\gamma^{\tau\tau}A_{\tau}^{(2)}+\gamma^{\tau\sigma}A_{\s}^{(2)}\,
. \eea
Substituting here the solution for the metric,
eqs.(\ref{metric1}), (\ref{metric2}), we  obtain remarkably simple
formulae
\bea
\gamma_{\tau\b}\epsilon^{\beta\rho}A_{\rho}^{(2)}&=&
\frac{i}{8}\Omega_{\tau}~{\bf J}\, ,\\
\gamma_{\sigma\b}\epsilon^{\beta\rho}A_{\rho}^{(2)}&=&
\frac{i}{4\sqrt{\lambda}}(J+{\cal
H})~{\bf J}\, ,
 \eea
where
$$
{\mathcal
H}=J-i\frac{\sqrt{\lambda}}{2}(\bar{\psi}\rho^1\psi'-
\bar{\psi}'\rho^1\psi)-J\bar{\psi}\psi
$$
is the Hamiltonian obtained from the Lagrangian (\ref{quadms}).

By using the equations of motion following from (\ref{quadms}) one
can prove the following on-shell relation \bea
\Omega_{\tau}=-i\frac{\sqrt{\lambda}}{J}\zeta_{\sigma}\, . \eea
Thus, we finally get \bea
\gamma_{\tau\b}\epsilon^{\beta\rho}A_{\rho}^{(2)}&=&
\frac{\sqrt{\lambda}}{8J}\zeta_{\sigma}~{\bf J}\, ,\\
\gamma_{\sigma\b}\epsilon^{\beta\rho}A_{\rho}^{(2)}&=&
\frac{i}{4\sqrt{\lambda}}(J+{\cal
H})~{\bf J}\, .
 \eea
In this way we completely excluded the metric in favor of
dynamical variables from the Lax representation.

\medskip

Now putting all the pieces of the Lax connection together we check
that the zero-curvature condition (\ref{zc}) is indeed satisfied
as the consequence of the dynamical equations for fermions derived
from the Lagrangian (\ref{quadms}). This proves that the model of
two-dimensional Dirac fermions defined by the Lagrangian
(\ref{quadms}) is integrable. Eigenvalues of the monodromy matrix
\bea {\rm T}(x)=\P {\rm exp} {\int_0^{2\pi}{\rm d
}\sigma\L_{\sigma}} \, \eea are the integrals of motion. Finally
we note that to get a connection with the Lagrangian (\ref{rf})
one has to rescale the fermion as $\psi\to
\frac{1}{\sqrt{J}}\psi$.

\section*{Acknowledgments}
We are grateful to V.~Bazhanov, N.~Beisert, M.~Staudacher,
A.~Tseytlin, M.~Zamaklar and K.~Zarembo for interesting
discussions. We would like to thank V. Kazakov  and all
 the organizers of the {\it Ecole Normale Sup\'erior
Summer Institute} in August 2005 where this work was completed for
an inspiring conference, and their warm hospitality. The work of
G.~A. was supported in part by the European Commission RTN
programme HPRN-CT-2000-00131 and by RFBI grant N02-01-00695. The
work of S.~F.~was supported in part by the EU-RTN network {\it Constituents, Fundamental
Forces and Symmetries of the Universe} (MRTN-CT-2004-005104).

\appendix

\section{Gamma-matrices}
Introduce the following five $4 \times 4$ matrices
\begin{eqnarray}
\nonumber \gamma_1&=&{\footnotesize\left(
\begin{array}{cccc}
  0 & 0 & 0 & 1 \\
  0 & 0 & -1 & 0 \\
   0 & -1 & 0 & 0 \\
   1 & 0 & 0 & 0
\end{array} \right)},\hspace{0.3in}\gamma_2={\footnotesize\left(
\begin{array}{cccc}
  0 & 0 & i & 0 \\
  0 & 0 & 0 & -i \\
   -i & 0 & 0 & 0 \\
   0 & i & 0 & 0
\end{array} \right)},\hspace{0.3in}\gamma_3={\footnotesize\left(
\begin{array}{cccc}
  0 & 0 & 1 & 0 \\
  0 & 0 & 0 & 1 \\
   1 & 0 & 0 & 0 \\
   0 & 1 & 0 & 0
\end{array} \right)}, \\
\nonumber \gamma_4&=&{\footnotesize \left(
\begin{array}{cccc}
  0 & 0 & 0 & -i \\
  0 & 0 & -i & 0 \\
   0 & i & 0 & 0 \\
   i & 0 & 0 & 0
\end{array} \right)},\hspace{0.28in}~\gamma_5={\footnotesize \left(
\begin{array}{cccc}
  i & 0 & 0 & 0 \\
  0 & i & 0 & 0 \\
   0 & 0 & -i & 0 \\
   0 & 0 & 0 & -i
\end{array} \right)}=-i\gamma_1\gamma_2\gamma_3\gamma_4 \, .
\end{eqnarray}
These matrices satisfy the SO(4,1) Clifford algebra
$$
\gamma_a\gamma_b+\gamma_b\gamma_a=2\eta_{ab}\, , ~~~~~a=1,\ldots,
5.
$$
where $\eta=\mbox{diag}(1,1,1,1,-1)$. Further, the matrices
$\gamma_a$ belong to the Lie algebra $\su(2,2)$ as they satisfy
the relation \bea \Sigma\gamma_a+\gamma_a^{\dagger}\Sigma=0\, ,
~~~~~~~~ \Sigma={\rm diag}(1,1,-1,-1). \eea

Analogously, the SO(5) Dirac matrices are
\begin{eqnarray}
\nonumber \Gamma_1&=&{\footnotesize\left(
\begin{array}{cccc}
  0 & 0 & 0 & -1 \\
  0 & 0 & 1 & 0 \\
   0 & 1 & 0 & 0 \\
   -1 & 0 & 0 & 0
\end{array} \right)},\hspace{0.3in}\Gamma_2={\footnotesize\left(
\begin{array}{cccc}
  0 & 0 & -i & 0 \\
  0 & 0 & 0 & i \\
   i & 0 & 0 & 0 \\
   0 & -i & 0 & 0
\end{array} \right)},\hspace{0.3in}\Gamma_3={\footnotesize\left(
\begin{array}{cccc}
  0 & 0 & -1 & 0 \\
  0 & 0 & 0 & -1 \\
   -1 & 0 & 0 & 0 \\
   0 & -1 & 0 & 0
\end{array} \right)}, \\
\nonumber \Gamma_4&=&{\footnotesize \left(
\begin{array}{cccc}
  0 & 0 & 0 & i \\
  0 & 0 & i & 0 \\
   0 & -i & 0 & 0 \\
   -i & 0 & 0 & 0
\end{array} \right)},\hspace{0.28in}~\Gamma_5={\footnotesize \left(
\begin{array}{cccc}
  1 & 0 & 0 & 0 \\
  0 & 1 & 0 & 0 \\
   0 & 0 & -1 & 0 \\
   0 & 0 & 0 & -1
\end{array} \right)}\, .
\end{eqnarray}
They satisfy the SO(5) Clifford algebra
$$
\Gamma_a\Gamma_b+\Gamma_b\Gamma_a=2\delta_{ab}\, .
$$
Moreover, all of them are hermitian, so that $i\Gamma_{a}$ belongs
to $\su(4)$.

\medskip

We represent the generators of the superconformal group by the
$\su(2,2)$ matrices. In particular, the generator of scaling
transformations is chosen to be
\medskip
\begin{eqnarray}
{\rm D}={\scriptsize \frac{1}{2}\left(
\begin{array}{cccc}
  1 & 0 & 0 & 0 \\
  0 & 1 & 0 & 0 \\
   0 & 0 & -1 & 0 \\
   0 & 0 & 0 & -1
\end{array} \right)\, }= -\ihalf\gamma_5  =\half \Gamma_5\, .
\end{eqnarray}
The generators of translations are given by
\medskip
\begin{eqnarray}
\nonumber {\rm P}^0={\scriptsize \left(
\begin{array}{cccc}
  0 & 0 & 0 & 0 \\
  0 & 0 & 0 & 0 \\
   i & 0 & 0 & 0 \\
   0 & i & 0 & 0
\end{array} \right)},\hspace{0.3in}{\rm P}^1={\scriptsize \left(
\begin{array}{cccc}
  0 & 0 & 0 & 0 \\
  0 & 0 & 0 & 0 \\
   i & 0 & 0 & 0 \\
   0 & -i & 0 & 0
\end{array} \right)},\hspace{0.3in}
{\rm P}^2={\scriptsize \left(
\begin{array}{cccc}
  0 & 0 & 0 & 0 \\
  0 & 0 & 0 & 0 \\
   0 & i & 0 & 0 \\
   i & 0 & 0 & 0
\end{array} \right)},\hspace{0.3in}{\rm P}^3={\scriptsize \left(
\begin{array}{cccc}
  0 & 0 & 0 & 0 \\
  0 & 0 & 0 & 0 \\
   0 & 1 & 0 & 0 \\
   -1 & 0 & 0 & 0
\end{array} \right)}
\end{eqnarray}
The conformal boosts are defined as \bea {\rm K}^i=({\rm
P}^i)^t,~~~~\mbox{for}~~i=0,3;\hspace{0.3in}{\rm K}^i=-({\rm
P}^i)^t,~~~~\mbox{for}~~i=1,2 \, .\eea
\noindent We also have \bea
{\rm P}^0+{\rm K}^0=-\gamma_3\, , ~~~~~~~~ {\rm
P}^3+{\rm K}^3=-\gamma_1\, , \\
{\rm P}^1+{\rm K}^1=-\gamma_2\, , ~~~~~~~~{\rm P}^2+{\rm
K}^2=-\gamma_4\, . \eea

\section{Global symmetry transformations}

\vskip 0.5cm \noindent{\sl Conformal Transformations} \vskip 0.5cm

If we
parametrize the $\su(2,2)$ matrix $\Phi$ parametrizing infinitezimal conformal transformation as
\bea
\Phi={\footnotesize
\left(\begin{array}{rrrr}
i\xi_1 ~&~ \alpha_1+i\beta_1 ~&~  \alpha_2+i\beta_2 ~&~  \alpha_3+i\beta_3 \\
- \alpha_1+i\beta_1 ~&~ i\xi_2 ~&~ \alpha_4+i\beta_4 ~&~  \alpha_5+i\beta_5 \\
 \alpha_2-i\beta_2 ~&~  \alpha_4-i\beta_4 ~&~ i\xi_3 ~&~  \alpha_6+i\beta_6 \\
 \alpha_3-i\beta_3 ~&~  \alpha_5-i\beta_5 ~&~ -
\alpha_6+i\beta_6 & -i(\xi_1+\xi_2+\xi_3)
\end{array}
\right)} \, .
\eea
then eq.(\ref{trga}) implies the following
transformation rules for the coordinates $v$:
\bea \nonumber
\delta v_1 &=&
(\beta_1+\beta_6)v_2+( \alpha_1- \alpha_6)v_3+(\xi_1+\xi_3)v_4+(\beta_3-\beta_4)v_5
+( \alpha_3- \alpha_4)v_6\\
\nonumber
\delta v_2 &=& -(\beta_1+\beta_6)v_1-(\xi_2+\xi_3)v_3-( \alpha_1+ \alpha_6)v_4
+(- \alpha_2+ \alpha_5)v_5+(\beta_2-\beta_5)v_6 \\
\nonumber \delta v_3 &=& (- \alpha_1+ \alpha_6)v_1+(\xi_2+\xi_3)v_2+(\beta_1-\beta_6)v_4
+(\beta_2+\beta_5)v_5+( \alpha_2+ \alpha_5)v_6 \\
\nonumber \delta v_4 &=&-(\xi_1+\xi_3) v_1
+( \alpha_1+ \alpha_6)v_2+(-\beta_1+\beta_6)v_3+( \alpha_3+ \alpha_4)v_5-(\beta_3+\beta_4)v_6\\
\nonumber \delta v_5 &=& (\beta_3-\beta_4)v_1+(- \alpha_2+
\alpha_5)v_2 +(\beta_2+\beta_5)v_3
+( \alpha_3+ \alpha_4)v_4+(\xi_1+\xi_2)v_6\\
\nonumber \delta v_6 &=& ( \alpha_3-
\alpha_4)v_1+(\beta_2-\beta_5)v_2 +( \alpha_2+ \alpha_5)v_3
-(\beta_3+\beta_4)v_4-(\xi_1+\xi_2)v_5\, .
 \eea

\vskip 0.5cm \noindent{\sl R-symmetry Transformations} \vskip 0.5cm

If we
parametrize the $\su(4)$ matrix $\Phi$ parametrizing infinitezimal R-symmetry transformation as
\bea
\Phi_{\su(4)}={\footnotesize
\left(\begin{array}{rrrr}
i\xi_1 ~&~ \alpha_1+i\beta_1 ~&~ \alpha_2+i\beta_2 ~&~ \alpha_3+i\beta_3 \\
-\alpha_1+i\beta_1 ~&~ i\xi_2 ~&~ \alpha_4+i\beta_4 ~&~ \alpha_5+i\beta_5 \\
-\alpha_2+i\beta_2 ~&~ -\alpha_4+i\beta_4 ~&~ i\xi_3 ~&~ \alpha_6+i\beta_6 \\
-\alpha_3+i\beta_3 ~&~ -\alpha_5+i\beta_5 ~&~ -\alpha_6+i\beta_6 &
-i(\xi_1+\xi_2+\xi_3)
\end{array}
\right)} \, .
\eea
then we find the following transformation rules for
the coordinates $u_i$:
\bea
\nonumber \delta u_1 &=&
(\beta_1+\beta_6)u_2+(\alpha_1-\alpha_6)u_3+(\xi_1+\xi_3)u_4+
(\alpha_3-\alpha_4)u_5+(\beta_4-\beta_3)u_6\\
\nonumber
\delta u_2 &=& -(\beta_1+\beta_6)u_1-(\xi_2+\xi_3)u_3-
(\alpha_1+\alpha_6)u_4
+(\beta_2-\beta_5)u_5+(\alpha_2-\alpha_5)u_6 \\
\nonumber \delta u_3 &=&
(-\alpha_1+\alpha_6)u_1+(\xi_2+\xi_3)u_2+(\beta_1-\beta_6)u_4
+(\alpha_2+\alpha_5)u_5-(\beta_2+\beta_5)u_6 \\
\nonumber \delta u_4 &=&-(\xi_1+\xi_3) u_1
+(\alpha_1+\alpha_6)u_2+(-\beta_1+\beta_6)u_3-(\beta_3+\beta_4)u_5-
(\alpha_3+\alpha_4)u_6
\\
\nonumber \delta u_5 &=&(-\alpha_3+\alpha_4) u_1
+(-\beta_2+\beta_5)u_2-(\alpha_2+\alpha_5)u_3+
(\beta_3+\beta_4)u_4+(\xi_1+\xi_2)u_6
\\
\nonumber \delta u_6 &=&
(\beta_3-\beta_4)u_1+(-\alpha_2+\alpha_5)u_2 +
(\beta_2+\beta_5)u_3 +(\alpha_3+\alpha_4)u_4-(\xi_1+\xi_2)u_5\, .
 \eea
Clearly, these transformations obey the constraint $u_i\delta
u_i=0$.

\section{Fermionic identities and conjugation rules}
Here we collect some formulas involving fermionic expressions.
Introducing the Dirac fermion $\psi$, see eq.(\ref{wsf}), we find
that \bea
\zeta_{\tau}&=&\bar{\psi}\rho^0\pa_{0}\psi-\pa_{0}\bar{\psi}\rho^0\psi
\, ,
\\
\Omega_{\sigma}&=&-i\left(\bar{\psi}\rho^1\pa_{1}\psi-
\pa_{1}\bar{\psi}\rho^1\psi\right)\,
,
  \\
\Lambda&=&\bar{\psi}\psi \, .\eea

\medskip
We have a few important identities which allow us to simplify the
structure of the Lagrangian. They include \bea \nonumber
&&\epsilon^{\a\beta}\Big[-\bar{\psi}\psi~\pa_{\a}\bar{\psi}\rho^5
\pa_{\beta}\psi+ \pa_{\a}\bar{\psi}
\pa_{\beta}\psi~\bar{\psi}\rho^5\psi
+\bar{\psi}\pa_{\a}\psi~\bar{\psi}\rho^5\pa_{\beta}\psi
-\pa_{\a}\bar{\psi}\psi~\pa_{\b}\bar{\psi}\rho^5\psi\Big]=\\
\nonumber
&&\epsilon^{\a\beta}\Big[-\pa_{\a}\left(\bar{\psi}\psi~
\bar{\psi}\rho^5\pa_{\beta}\psi+\pa_{\b}\bar{\psi}\psi~
\bar{\psi}\rho^5\psi\right)
+2\left(\bar{\psi}\pa_{\a}\psi~\bar{\psi}\rho^5\pa_{\beta}
\psi-\pa_{\alpha}\bar{\psi}\psi~\pa_{\beta}\bar{\psi}\rho^5\psi\right)
\Big] \eea and
\bea
 \nonumber
&&\epsilon^{\a\beta}\Big[\bar{\psi}\psi\left(-\bar{\psi}\psi~
\pa_{\a}\bar{\psi}\rho^5
\pa_{\beta}\psi+ \pa_{\a}\bar{\psi}
\pa_{\beta}\psi~\bar{\psi}\rho^5\psi
+\bar{\psi}\pa_{\a}\psi~\bar{\psi}\rho^5\pa_{\beta}\psi
-\pa_{\a}\bar{\psi}\psi~\pa_{\b}\bar{\psi}\rho^5\psi\right)\Big]\\
&&=-(\bar{\psi}\psi)^2\epsilon^{\alpha\beta}\pa_{\a}\bar{\psi}\rho^5
\pa_{\beta}\psi\, . \eea

The following identity is valid \bea
\bar{\psi}\rho^0\pa_{0}\psi~\bar{\psi}\rho^1\pa_{1}\psi
=-\bar{\psi}\pa_0\psi~\bar{\psi}\rho^5\pa_1\psi \eea In addition
the properties of the $\rho$-matrices imply the following complex
conjugation rules: \bea
(\bar{\psi}\pa_{\a}\psi)^*=\pa_{\a}\bar{\psi}\psi,~~~~~~~~
(\bar{\psi}\rho^5\pa_{\a}\psi)^*=-\pa_{\a}\bar{\psi}\rho^5\psi\eea

\section{Change of variables}
We have found that in order to bring  the kinetic term
(\ref{quad1}) to the canonical form (\ref{quad2}) the following
non-linear shift of the fermions in eq.(\ref{quad1}) should be
performed (here $\kappa=\frac{\sqrt{\lambda'}}{2}$)
 \bea
 \nonumber
 \psi &\to&
\psi+\frac{i\kappa}{J^2}\rho^1\psi
~(\pa_1\bar{\psi}\psi)-\frac{i\kappa}{8J^3}\rho^1\pa_1\psi
~(\bar{\psi}\psi)^2 -\frac{\kappa^2}{4J^4}\pa_1^2\psi
~(\bar{\psi}\psi)^2 \\
\nonumber
&&~~-\frac{\kappa^2}{2J^4}\pa_1\psi~(\pa_1\bar{\psi}\psi-
\bar{\psi}\pa_1\psi)\bar{\psi}\psi
+\frac{\kappa^2}{2J^4}\rho^0\psi
~(\pa_1\bar{\psi}\rho^0\pa_1\psi)\bar{\psi}\psi \\
&&~~-\frac{5}{4}\frac{i\kappa^3}{J^6}\rho^1\pa_1\psi
(\pa_1\bar{\psi}\pa_1\psi) ~(\bar{\psi}\psi)^2\, , \nonumber
\\
\nonumber
\\
\nonumber
 \bar{\psi} &\to&
\bar{\psi}-\frac{i\kappa}{J^2}\bar{\psi}\rho^1~(\bar{\psi}\pa_1\psi)
+\frac{i\kappa}{8J^3}\pa_1\bar{\psi}\rho^1~(\bar{\psi}\psi)^2-
\frac{\kappa^2}{4J^4}\pa_1^2\bar{\psi}~(\bar{\psi}\psi)^2
\\
\nonumber
&&~~+\frac{\kappa^2}{2J^4}\pa_1\bar{\psi}~(\pa_1\bar{\psi}\psi-
\bar{\psi}\pa_1\psi)\bar{\psi}\psi
+\frac{\kappa^2}{2J^4}\bar{\psi}\rho^0
~(\pa_1\bar{\psi}\rho^0\pa_1\psi)\bar{\psi}\psi  \\
\label{directshift}
&&~~+\frac{5}{4}\frac{i\kappa^3}{J^6}\pa_1\bar{\psi}\rho^1
(\pa_1\bar{\psi}\pa_1\psi) ~(\bar{\psi}\psi)^2\, .\eea Under this
shift the Hamiltonian (\ref{simple}) transforms into expression
(\ref{complicated}). Below we also give the formulae for the
transformation inverse to (\ref{directshift}): \bea
 \nonumber
 \psi &\to&
\psi-\frac{i\kappa}{J^2}\rho^1\psi
~(\pa_1\bar{\psi}\psi)+\frac{i\kappa}{8J^3}\rho^1\pa_1\psi
~(\bar{\psi}\psi)^2 -\frac{\kappa^2}{4J^4}\pa_1^2\psi
~(\bar{\psi}\psi)^2 \\
\nonumber
&&~~-\frac{\kappa^2}{2J^4}\pa_1\psi~(\pa_1\bar{\psi}\psi+
\bar{\psi}\pa_1\psi)\bar{\psi}\psi
-\frac{\kappa^2}{2J^4}\rho^0\psi
~(\pa_1\bar{\psi}\rho^0\pa_1\psi)\bar{\psi}\psi \\
&&~~-\frac{i\kappa^3}{4J^6}\rho^1\pa_1\psi
(\pa_1\bar{\psi}\pa_1\psi) ~(\bar{\psi}\psi)^2\, , \nonumber
\\
\nonumber
\\
\nonumber
 \bar{\psi} &\to&
\bar{\psi}+\frac{i\kappa}{J^2}\bar{\psi}\rho^1~(\bar{\psi}\pa_1\psi)
-\frac{i\kappa}{8J^3}\pa_1\bar{\psi}\rho^1~(\bar{\psi}\psi)^2-
\frac{\kappa^2}{4J^4}\pa_1^2\bar{\psi}~(\bar{\psi}\psi)^2
\\
\nonumber
&&~~-\frac{\kappa^2}{2J^4}\pa_1\bar{\psi}~(\pa_1\bar{\psi}\psi+
\bar{\psi}\pa_1\psi)\bar{\psi}\psi
-\frac{\kappa^2}{2J^4}\bar{\psi}\rho^0
~(\pa_1\bar{\psi}\rho^0\pa_1\psi)\bar{\psi}\psi  \\
\label{inverseshift}
&&~~+\frac{i\kappa^3}{4J^6}\pa_1\bar{\psi}\rho^1
(\pa_1\bar{\psi}\pa_1\psi) ~(\bar{\psi}\psi)^2\, .\eea

\section{Poisson structure}
\bea
\{\psi_1,\psi_1\}=&-&\frac{i\sqrt{\lambda'}}{2J}(\psi_1\psi_2)'
\nonumber\\
\nonumber
&+&\frac{i\sqrt{\lambda'}}{4J^2}(\psi_1\psi_2)'(\psi_2\psi^*_2
-2\sqrt{\lambda'}\psi_2\psi'^*_1+
\sqrt{\lambda'}\psi^*_2\psi'_1 )\\
\nonumber
&+&\frac{i{\lambda'}^{\frac{3}{2}}}{8J^3}(\psi_1\psi_2)'^2(\psi^*_1\psi'^*_2
+3\psi^*_2\psi'^*_1)\, ,\\
\{\psi_2,\psi_2\}=&-&\frac{i\sqrt{\lambda'}}{2J}(\psi_1\psi_2)'
\nonumber\\
&-&\frac{i\sqrt{\lambda'}}{4J^2}(\psi_1\psi_2)'(\psi_1\psi^*_1
-2\sqrt{\lambda'}\psi_1\psi'^*_2+
\sqrt{\lambda'}\psi^*_1\psi'_2 )
\nonumber \\
\nonumber
&-&\frac{i{\lambda'}^{\frac{3}{2}}}{8J^3}(\psi_1\psi_2)'^2(3\psi^*_1\psi'^*_2
+\psi^*_2\psi'^*_1)\, ,\\
\nonumber
\{\psi_1,\psi_2\}=&&\frac{i\sqrt{\lambda'}}{8J^2}(\psi_1\psi_2)'
(\psi_1\psi^*_2+\psi_1^*\psi_2+2\sqrt{\lambda'}
(\psi_1\psi^*_1-\psi_2\psi^*_2)')\\
\nonumber
&-&\frac{i\lambda'}{4J^3}(\psi_1\psi_2)'^2(-\psi^*_1\psi^*_2
+\sqrt{\lambda'}(\psi_1^*\psi'^*_1-\psi^*_2\psi'^*_2))\, ,
 \eea

\bea \nonumber \{\psi_1,\psi^*_2\}=&-&\frac{i\sqrt{\lambda'}}{2J}
(\psi_1\psi'^*_1+\psi^*_2\psi'_2)\\
\nonumber &+&\frac{i\sqrt{\lambda'}}{8J^2}
\big[(-\psi_1\psi_2\psi^*_1\psi'^*_2+\psi_2\psi^*_1\psi^*_2\psi'_1
+2\psi_1\psi_2\psi^*_2\psi'^*_1-2\psi_1\psi^*_1\psi^*_2\psi'_2
)\\
\nonumber &&~~~~~~~~+ 2\sqrt{\lambda'}(
-\psi_1\psi^*_1\psi'_2\psi'^*_1
+\psi_1\psi^*_2\psi'_1\psi'^*_1+\psi_1\psi^*_2\psi'_2\psi'^*_2
-\psi_2\psi^*_2\psi'_2\psi'^*_1)
\big]\\
\nonumber &-& \frac{i\lambda'^{\frac{3}{2}}}{4J^3}
\big[\psi_1\psi_2\psi^*_1\psi'_1\psi'^*_1\psi'^*_2+
\psi_2\psi^*_1\psi^*_2\psi'_1\psi'_2\psi'^*_2\big]\, ,\eea

\bea \nonumber \{\psi_2,\psi^*_1\}=&&\frac{i\sqrt{\lambda'}}{2J}
(\psi^*_1\psi'_1+\psi_2\psi'^*_2)\\
\nonumber &+&\frac{i\sqrt{\lambda'}}{8J^2}
\big[(\psi_1\psi_2\psi^*_2\psi'^*_1-\psi_1\psi^*_1\psi^*_2\psi'_2
-2\psi_1\psi_2\psi^*_1\psi'^*_2+2\psi_2\psi^*_1\psi^*_2\psi'_1
)\\
\nonumber &&~~~~~~~~+ 2\sqrt{\lambda'}(
-\psi_1\psi^*_1\psi'_1\psi'^*_2
+\psi_2\psi^*_1\psi'_1\psi'^*_1+\psi_2\psi^*_1\psi'_2\psi'^*_2
-\psi_2\psi^*_2\psi'_1\psi'^*_2)
\big]\\
\nonumber &+& \frac{i\lambda'^{\frac{3}{2}}}{4J^3}
\big[\psi_1\psi_2\psi^*_2\psi'_2\psi'^*_1\psi'^*_2+
\psi_1\psi^*_1\psi^*_2\psi'_1\psi'_2\psi'^*_1\big] \, .\eea

\bea \nonumber
\{\psi_1,\psi^*_1\}=i&+&\frac{i\sqrt{\lambda'}}{2J}
(\psi_2\psi'^*_1-\psi^*_2\psi'_1)\\
\nonumber \nonumber
&+&\frac{i\sqrt{\lambda'}}{8J^2}\big[\psi_2\psi^*_2
(\psi_1\psi'^*_2-\psi^*_1\psi'_2)\\
\nonumber &&~~~~~~~+ 2\sqrt{\lambda'}(
\psi_1\psi_2\psi'^*_1\psi'^*_2-
\psi_1\psi^*_1\psi'_1\psi'^*_1+\psi_1\psi^*_1\psi'_2\psi'^*_2
-\psi_1\psi^*_2\psi'_2\psi'^*_1\\
\nonumber
&&~~~~~~~~~~~~~~~~~-\psi_2\psi^*_1\psi'_1\psi'_2-\psi_2\psi^*_2\psi'_1\psi'^*_1
-\psi_2\psi^*_2\psi'_2\psi'^*_2+ \psi_1^*\psi^*_2\psi'_1\psi'_2)
\big]\\
\nonumber &+&
\frac{i\lambda'}{4J^3}\big[\psi_1\psi_2\psi^*_1\psi^*_2(-\psi'_1\psi'^*_1
+2\psi'_2\psi'^*_2)\\
\nonumber &&~~~~~~+\sqrt{\lambda'}(-
\psi_1\psi_2\psi^*_2\psi'_1\psi'^*_1\psi'^*_2
 +
\psi_2\psi^*_1\psi^*_2\psi'_1\psi'_2\psi'^*_1\\
\nonumber
&&~~~~~~~~~~~~~~~~~~~~~~~~~~~~~+2\psi_1\psi_2\psi^*_1\psi'_2\psi'^*_1\psi'^*_2
-2\psi_1\psi^*_1\psi^*_2\psi'_1\psi'_2\psi'^*_2)\big]\\
\nonumber &-&
\frac{3i\lambda'^2}{4J^4}\psi_1\psi_2\psi^*_1
\psi^*_2\psi'_1\psi'_2\psi'^*_1\psi'^*_2\,
,\eea

\bea \nonumber
\{\psi_2,\psi^*_2\}=i&-&\frac{i\sqrt{\lambda'}}{2J}
(\psi_1\psi'^*_2-\psi^*_1\psi'_2)\\
\nonumber
&+&\frac{i\sqrt{\lambda'}}{8J^2}\big[(\psi_1\psi^*_1
(\psi_2\psi'^*_1-\psi^*_2\psi'_1)\\
\nonumber &&~~~~~~~~+ 2\sqrt{\lambda'}(
\psi_1\psi_2\psi'^*_1\psi'^*_2-
\psi_1\psi^*_1\psi'_1\psi'^*_1-\psi_1\psi^*_1\psi'_2\psi'^*_2
-\psi_1\psi^*_2\psi'_2\psi'^*_1\\
\nonumber
&&~~~~~~~~~~~~~~~~~~~~-\psi_2\psi^*_1\psi'_1\psi'_2+
\psi_2\psi^*_2\psi'_1\psi'^*_1
-\psi_2\psi^*_2\psi'_2\psi'^*_2+ \psi_1^*\psi^*_2\psi'_1\psi'_2)
\big]\\
\nonumber &+&
\frac{i\lambda'}{4J^3}\big[\psi_1\psi_2\psi^*_1\psi^*_2(-2\psi'_1\psi'^*_1
+\psi'_2\psi'^*_2)\\
\nonumber
&&~~~~~~~~+\sqrt{\lambda'}(\psi_1\psi_2\psi^*_1\psi'_2\psi'^*_1\psi'^*_2
-\psi_1\psi^*_1\psi^*_2\psi'_1\psi'_2\psi'^*_2\\
\nonumber &&~~~~~~~~~~~~~~~~~~~~~~~~~~~~~-
2\psi_1\psi_2\psi^*_2\psi'_1\psi'^*_1\psi'^*_2
 +
2\psi_2\psi^*_1\psi^*_2\psi'_1\psi'_2\psi'^*_1 )\big]\\
\nonumber &-& \frac{3i\lambda'^2}{4J^4}\psi_1\psi_2\psi^*_1
\psi^*_2\psi'_1\psi'_2\psi'^*_1\psi'^*_2\, .\eea The Poisson
bracket is ultra-local. Here for the sake of simplicity we have
omitted on the r.h.s. the overall delta function
$\delta(\sigma-\sigma')$.


\end{document}